\documentclass[aps,amssymb,preprint,tightenlines]{revtex4}
\usepackage{graphicx,psfrag}

\renewcommand{\a}{\alpha}
\renewcommand{\b}{\beta}

\renewcommand{\d}{\delta}\newcommand{\D}{\Delta}
\newcommand{\e}{\epsilon}

\renewcommand{\l}{\lambda}\renewcommand{\L}{\Lambda}
\renewcommand{\t}{\theta}
\renewcommand{\tt}{{\tilde{\theta}}}
\newcommand{\ttt}{\,{\tilde{\!\tilde{\theta}}}}
\newcommand{\ttG}{\,{\tilde{\!\tilde{G}}}}
\newcommand{\T}{\Theta}
\newcommand{\n}{\eta}
\newcommand{\s}{\sigma}

\renewcommand{\r}{\rho}
\renewcommand{\u}{\mu}
\renewcommand{\v}{\nu}
\newcommand{\vn}{\vec{n}}
\newcommand{\vu}{\vec{u}}

\newcommand{\cI}{{\mathcal I}}

\newcommand{\cT}{{\mathcal T}}
\newcommand{\cC}{{\mathcal C}}
\newcommand{\cD}{{\mathcal D}}
\newcommand{\tcD}{{\tilde{\mathcal D}}}

\newcommand{\real}{\mathbb{R}}
\newcommand{\comp}{\mathbb{C}}
\newcommand{\Nat}{\mathbb{N}}
\newcommand{\Z}{\mathbb{Z}}
\newcommand{\SO}{\mathrm{SO}}
\newcommand{\SU}{\mathrm{SU}}
\newcommand{\SL}{\mathrm{SL}}

\newcommand{\be}{\begin{equation}}
\newcommand{\ee}{\end{equation}}
\newcommand{\beq}{\begin{eqnarray}}
\newcommand{\eeq}{\end{eqnarray}}

\renewcommand{\det}{\mathrm{det}}

\newcommand{\w}{\wedge}

\newcommand{\tr}{\mathrm{tr}}

\newtheorem{theo}{Theorem}
\newtheorem{prop}{Proposition}

\newcommand{\proof}[1]{\vspace{5mm} \noindent \textit{Proof : ---} \\
{#1} \begin{flushright}$\blacksquare$\end{flushright}}

\begin{document}
\begin{titlepage}
\title{\large Asymptotics of 6j and 10j symbols}
\author{ Laurent Freidel}
\email{lfreidel@perimeterinstitute.ca}
\affiliation{\vspace{2mm}Perimeter Institute for Theoretical
Physics\\ 35 King street North, Waterloo  N2J-2G9,Ontario,
Canada\\} \affiliation{Laboratoire de Physique, \'Ecole Normale
Sup{\'e}rieure de Lyon \\ 46 all{\'e}e d'Italie, 69364 Lyon Cedex
07, France }

\author{David Louapre}
\email{dlouapre@ens-lyon.fr}
\affiliation{\vspace{2mm}Laboratoire de Physique, \'Ecole Normale
Sup{\'e}rieure de Lyon \\ 46 all{\'e}e d'Italie, 69364 Lyon Cedex
07, France}\thanks{UMR 5672 du CNRS}


\begin{abstract}

It is well known that the building blocks for state sum models of
quantum gravity are given by $6j$ and $10j$ symbols. In this work
we study the asymptotics of these symbols by using their
expressions  as group integrals. We carefully describe the measure
involved in terms of invariant variables and develop new technics
in order to study their asymptotics. Using these technics we
compute the asymptotics of the various Euclidean and Lorentzian $6j$-symbols.
 Finally we compute the
asymptotic expansion of the $10j$ symbol which is shown to be
non-oscillating, in agreement with a recent result of Baez et al.
We discuss the physical origin of this behavior and a way to
modify the Barrett-Crane model in order to cure this disease.
\end{abstract}
\maketitle
\end{titlepage}

\section{Introduction}
It is well known that state sum models defined over a
triangulation of space time are a central tool in the construction
of discrete transition amplitudes for quantum gravity. It was
first understood by Ponzano and Regge \cite{PR} that the
transition amplitude of 3d Euclidean gravity could be expressed as
a state sum model defined over a triangulation of space time. This
state sum model is  obtained by summing, over $\SU(2)$
representations labelling the edges of the triangulation, a given
weight depending on this labelling. This sum over representations
is interpreted as a sum over geometry, and the weight is a product
of $\SU(2)$ $6j$ symbols associated with each colored tetrahedron.
This construction can be extended to the case of 4d gravity and
$2+1$ Lorentzian gravity. It was shown in \cite{Flor,davids} that
the building block in the construction of transition amplitudes
for $2+1$ Lorentzian gravity are the  $\SL(2,\real)$ $6j$ symbols.
In the context of 4 dimensional gravity, Barrett and Crane
\cite{BarC} proposed to use the so-called $10j$ symbol in order to
construct the discretized transition amplitudes.

One of the key argument in favor of the Ponzano-Regge model as a
model for 3d gravity is the fact that the asymptotic behavior of
the $6j$ symbol reproduces the discretized Regge action for 3d
gravity. This asymptotic has been conjectured in 1968 by Ponzano
and Regge but proved only in 1999 by Roberts \cite{Roberts}. In
the same way, one of the argument in favor of the Barrett-Crane
amplitude as a building block for quantum gravity was the fact
shown by Barrett and Williams \cite{BarW} that the stationary
contribution to the asymptotic of the $10j$ symbol reproduces also
the discretized Regge action of 4d gravity. However, recent
numerical simulations and computations by Baez et al.
\cite{Baezpos, Baez1} have shown that the asymptotic of the 10j
symbol is not dominated by an oscillating contribution.

In this paper we address this issues by using expressions for the
$6j$ and $10j$ symbol as integrals over group elements
\cite{Barint}. This is done by carefully describing the measure
involved in terms of gauge invariant variables. We show that the
asymptotics separates into a contribution coming from a stationary
phase approximation and a contribution coming from degenerates
configurations associated with singularities of the integrand. The
stationary phase produces an oscillatory Regge behavior, whereas
the degenerate contribution is non oscillating. It should be noted
that this general method has been outlined in the recent paper of
Baez, Christensen and Egan \cite{Baez1}, and even if our approach
is independent from their, one can view our paper as giving proofs
of the conjectures they made. Also, while we were completing the
redaction of this work, we become aware of the very recent article
\cite{Baras}, which is leading to the same conclusions as ours but
using a different approach.

In section \ref{sec:as} we express the square of the $\SU(2)$
$6j$-symbol as an integral over the space of spherical tetrahedra.
We show that the measure of integration is given by the inverse
square root of the determinant of the Graham matrix \cite{alek}
associated with spherical tetrahedra. We show that the integral
naturally separates into two parts : one for which the asymptotics
is dominated by an oscillating contribution, associated with flat
non-degenerate tetrahedra and obtained by a stationary phase
approximation; and one for which the asymptotics comes from
boundary contributions which label degenerate tetrahedra. We prove
that this contribution is given by an integral associated with the
Euclidean group, as was conjectured by Baez et al \cite{Baez1}.

In section \ref{sec:lorentz} we express the square of the
$\SL(2,\real)$ $6j$-symbol as an integral over the space of AdS
tetrahedra. We also split this integral in two parts, and show
that the stationary phase is dominated by an oscillating
contribution associated with flat non-degenerate Lorentzian
tetrahedra. The phase of this contribution is the Lorentzian Regge
action and the module is the volume of the Lorentzian tetrahedron.
In section \ref{sec:10j}  we apply our technics to the
$10j$-symbol and show that the leading contribution is a
non-oscillating one, dominating the oscillating Regge action term
found by Barrett and Williams. Finally  in section \ref{sec:disc}
we discuss the physical meaning and origin of this non-oscillating
behavior for the $10j$-symbol, in the spirit of the statistical
mechanics models of 'order by disorder', and we propose a
modification of the Barrett-Crane model to avoid this problem and
recover an oscillatory Regge action behavior.

\section{Asymptotics of the 6j-symbol}\label{sec:as}

\subsection{Integral expression for the square of the 6j-symbol}

We are interested in the computation of the asymptotics of the
square of the $6j$-symbol. The $6j$-symbol is a real number which
is associated to the labelling of the edges of a tetrahedra by
$SU(2)$ representations. It is obtain by combining four normalized
Clebsch-Gordan coefficients along the six edges of a tetrahedra
(see \cite{Roberts}). Let us denote by $V_l$ the $SU(2)$
representation of spin $l/2$ and by $\chi_l(g)=\tr_{V_l}(g)$ the
character of this representation. Lets $I=0,1,2,3$ be the 4
vertices of a tetrahedra $T$ and $(IJ)$ the edges of $T$. We
associate a representation $V_{l_{IJ}}$ to each edge $(IJ)$ of
$T$. It is well known \cite{Barint,FK} that the square of the
$6j$-symbol can be expressed as the following Feynman integral
over $\SU(2)$
\begin{equation}\label{6j}
I(l_{IJ})= \left\{\begin{array}{ccc}
l_{01} & l_{02} & l_{03}\\
l_{23}& l_{13} & l _{12}
\end{array}\right\}^2
= \int_{G^4}\  [dg_I] \prod_{I<J} \chi_{l_{IJ}}(g_J g_I^{-1}),
\label{eqn:Icoset}
\end{equation}
where the normalized Haar measure $dg$ is used. Such an identity
is clear since the integral over each group element produces a
pair of Clebsch-Gordan coefficient which are then combined into
one tetrahedron and its mirror image. It is also clear from this
expression that the $I(l_{IJ})\neq 0$ only if the $|l_{IK}-
l_{IL}|\leq l_{IJ} \leq l_{IK}+l_{IL}$ and $l_{IK}+l_{IL}+l_{IJ}$
is an even integer. In this expression, the integral is over four
copies of the group,
 however the integrand is invariant under the transformations
 \begin{equation}\label{inv}
 g_I\rightarrow h k g_I k^{-1},
 \end{equation}
 where $h,k$ are $SU(2)$ elements. It is therefore possible to
 gauge out this symmetry and write the integral purely in term of
 gauge invariant variables.
 A natural choice for the gauge invariant variables is given by the
 6 angles $\t_{IJ} \in [0,\pi]$:
 \begin{equation}\label{angle}
 \cos\t_{IJ} = {1\over 2} tr_{V_1}(g_J g_I^{-1}).
 \end{equation}
Note that this gauge symmetry has a geometrical interpretation.
The four group elements $g_I$ define 4 points, hence a
tetrahedron, in $S^3$. The angles $\t_{IJ}$ are the spherical
lengths of the edges of this tetrahedron, which indeed
parameterize its invariant geometry. Notice that the symmetry
includes reflexions $g\to -g$.
 The integral (\ref{eqn:Icoset}) can be written in terms of
these angles as follows:
\begin{theo}\label{theo1}
 \be \label{int:eucl}
 I(l_{IJ})=
 \frac{2}{\pi^4}
\int_{\cD_{\pi}} [\prod d\t_{IJ}] \frac{\prod_{I<J}
\sin((l_{IJ}+1)\t_{IJ})}{\sqrt{\det[\cos\t_{IJ}]}}
\ee where
$\cD_{\pi}$ is the subset of $[0;\pi]^6$ of angles satisfying the
relations :
\begin{eqnarray}
\t_{IJ} &\leq& \t_{IK}+\t_{JK}, \\
 2\pi & \geq &\t_{IJ}+\t_{IK}+\t_{JK},
\end{eqnarray}
for any triple $(I,J,K)$ of distinct elements. Geometrically this
domain is the set of all possible spherical tetrahedra.
\end{theo}
The denominator of this integral is the square root of the
determinant of the $4\times 4$ Graham matrix $G_{IJ}=\cos\t_{IJ}$
associated with the corresponding spherical tetrahedron . This
determinant is zero if and only if $\t_{IJ}$ belongs to the
boundary of the domain $\cD_{\pi}$, which is the set of degenerate
tetrahedra having zero volume. Note that if we consider $\t_{IJ}$
to be the dihedral angles of a spherical tetrahedra, $\partial
\cD_\pi$ is the set of flat tetrahedra (see the remark 3 below).

\begin{figure}[ht]
\psfrag{1}{$g_1$}\psfrag{2}{$g_2$}\psfrag{3}{$g_3$}\psfrag{4}{$g_0$}
\psfrag{t1}{$\t_{1}$}\psfrag{t2}{$\t_{2}$}\psfrag{t3}{$\t_{3}$}
\psfrag{t12}{$\t_{12}$}\psfrag{t13}{$\t_{13}$}\psfrag{t23}{$\t_{23}$}
\psfrag{tt12}{$\tt_{12}$}\psfrag{tt13}{$\tt_{13}$}\psfrag{tt23}{$\tt_{23}$}
\begin{center}
\includegraphics[width=9cm]{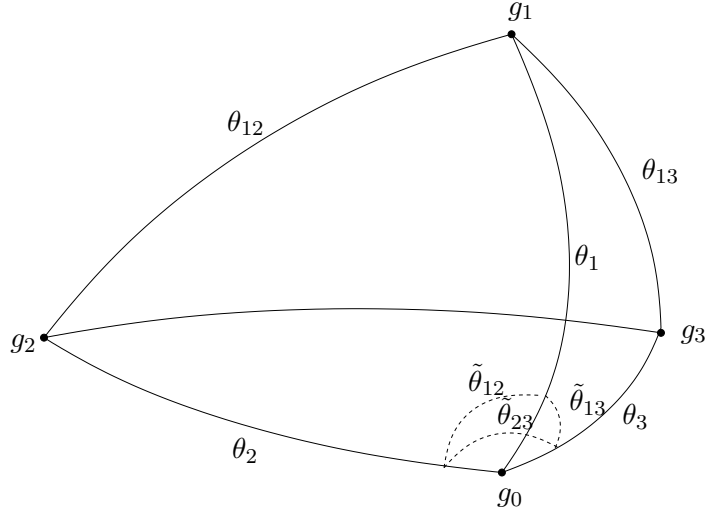}
\end{center}
\caption{Dual angles in a spherical tetrahedron. The dual
$\ttt_{23}$ of $\tt_{23}$ in the dashed spherical triangle appears
as the dihedral angle of the edge 01.}\label{tet}
\end{figure}

\proof{ \underline{Notations and spherical geometry}: The
invariant measure is obtained by a Faddev-Popov procedure. In
order to present the details of this procedure, we first need to
do a little bit of spherical geometry and introduce a new set of
angles denoted $\t_i,\tt_{ij},\ttt_{ij}\in[0,\pi]$, $i=1,2,3$.
This angles are related in the following way to the angle
$\t_{IJ}$ of eq.(\ref{angle}). \beq \t_i&=&\t_{0i}\\
\cos\t_{ij}&=&
\cos\t_{i}\cos\t_{j}+\sin\t_{i}\sin\t_{j}\cos\tt_{ij},
\label{rangle}\\ \cos\tt_{ij}&=&
\cos\tt_{ik}\cos\tt_{kj}+\sin\tt_{ik}\sin\tt_{kj}\cos\ttt_{ij},
\label{ttt} \eeq where in the eq.(\ref{ttt}), $i,j,k$ is any
permutation of $1,2,3$. The geometrical interpretation of these
angles is as follows : first $\t_i$ are the  lengths of the edges
$0i$, $\tt_{ij}$ is the angle between the edges $(0i)$ and $(0j)$,
finally $\ttt_{ij}$ is the dihedral angle of the edge $(0k)$
opposite to the edge $(ij)$ (see fig.\ref{tet}). In order to see
this, let $\sigma_i, i=1,2,3$ be the Pauli matrices,
$\sigma_i\sigma_j =\delta_{ij} +i \epsilon_{ij} ^k \sigma_k$ and
let us introduce the normalized vectors $\vn_i \in S^2$ such that
\be\label{exp}
 g_ig_0^{-1}=e^{i\theta_i \vec{n}_i\cdot\vec{\sigma}}\
 = \cos\t_i +i \sin\t_i (\vec{n}_i \cdot \vec{\sigma}).
\ee
Then
\beq
\cos\t_{ij}=\frac{1}{2} \tr(g_jg_i^{-1})= \frac{1}{2} \tr\left(g_jg_0^{-1}(g_ig_0^{-1})^{-1}\right)=
\cos\t_i\cos\t_{j}+\sin\t_{i}\sin\t_{j}(\vn_{i}\cdot \vn_{j}), \label{eqn:sphericaldual}
\eeq
so  $\cos\tt_{ij} =\vn_{i}\cdot \vn_{j}$.
Let us introduce the unit normal vector to the face $(0ij)$:
$\vec{a}_k = \vn_{i}\wedge \vn_{j}/|\vn_{i}\wedge \vn_{j}|$, where
 $ijk$ is a cyclic permutation of $123$.
A simple computation shows that \beq \cos\ttt_{ij} =-\vec{a}_i
\cdot \vec{a}_j, \eeq which characterizes $\ttt_{ij}$ as the
dihedral angle of the edge $0k$.

\noindent\underline{Measure:} Using these geometrical elements, we
can compute the invariant measure. In order to get this measure,
we first have to fix the symmetry by choosing a gauge. Using the
isometry (eq. \ref{inv}) we can first translate the tetrahedron so
that one of its vertex, say $g_0$, is at the identity. This fix
one $SU(2)$ invariance, the other corresponds to rotation of the
tetrahedron around the identity. We can gauge out part of this
invariance by fixing the direction of one edge say $(10)$. This
still let the freedom to rotate the tetrahedron around that edge,
we then choose to fix the direction of the plane $(120)$. In term
of the variables $\t_i,\vn_i$ the Haar measure reads $dg_i=
{2\over \pi} \sin^2\theta_i d\theta_i d^2\vec{n}_i$, where
$d^2\vec{n}$ is the normalized measure on the 2-sphere. So after
fixing $g_0 =1$ the measure $d\mu= dg_0 dg_1 dg_2 dg_3$ becomes
\be d\mu = \left({2\over \pi }\right)^3 \left(\prod_{i=1}^3
\sin^2\theta_i d\theta_i\right) d^2\vn_i. \ee The residual gauge
invariance is fixed by imposing \be \vec{n}_1= \left(
\begin{array}{c}
1 \\
0 \\
0
\end{array}
\right),\,
\vec{n}_2=
\left(
\begin{array}{c}
\cos {\tilde{\theta}}_{12} \\
\sin\tt_{12} \\
0
\end{array}
\right),\,
\vec{n}_3=
\left(
\begin{array}{c}
\cos\tt_{13} \\
\sin\tt_{13}\cos\ttt_{23} \\
\sin\tt_{13}\sin\ttt_{23}\\
\end{array}
\right) \ee In terms of this variables the measure is now \be d\mu
={2 \over \pi^4} \left(\prod_{i=1}^3 \sin^2\theta_i d\t_i \right)
\sin\tt_{12}d\tt_{12} \sin\tt_{13}d\tt_{13}\,  d\ttt_{23}. \ee It
can be checked directly that $\int d\mu =1$. Using relations
(\ref{ttt}) we can express the measure in terms of $\theta_i$ and
$\tt_{ij}$ \beq\label{mu1} d\mu ={2 \over \pi^4} \left(
\prod_{i=1}^3 \sin^2\theta_i d\t_i \right) { \prod_{i<j}
\sin\tt_{ij} d\tt_{ij}  \over |(\vec{n}_1 \wedge
\vec{n}_2)\cdot\vec{n}_3| }. \eeq This trivially follows from a
change of variable and the following fact $ |(\vec{n}_1 \wedge
\vec{n}_2)\cdot\vec{n}_3|= \sin\tt_{12}\sin\tt_{13}\sin\ttt_{23}$.
If we make the change of variable $\tt_{ij}\rightarrow \t_{ij}$,
eq.(\ref{mu1}) becomes \be\label{mu2} d\mu={2 \over \pi^4} {
\prod_{I<J}\sin\t_{IJ}d\theta_{IJ} \over \left(\prod_{i=1}^3
\sin\theta_i\right) |(\vec{n}_1 \wedge \vec{n}_2)\cdot\vec{n}_3|
}. \ee We can now show that the denominator is indeed
\be\label{sqrt} \sqrt{\det[\cos\t_{IJ}]} = \left(\prod_{i=1}^3
\sin\theta_i\right) |(\vec{n}_1 \wedge \vec{n}_2)\cdot\vec{n}_3|
\ee This is done by considering the $4\times 4$ determinant
$\det[\cos\t_{IJ}]$, which can be written as a $3\times 3 $
determinant $\det(\cos\t_{ij}-\cos\t_i \cos\t_j)$ by appropriate
line substraction. Moreover $\cos\t_{ij}-\cos\t_i \cos\t_j =
\sin\t_i \sin\t_j\cos\tt_{ij}$ so $\det[\cos\t_{IJ}]=
\left(\prod_{i=1}^3 \sin^2\theta_i  \right) det(\vn_i\cdot\vn_j)$.
If we call $n_i^a$ the component of $\vn_i$ in any given basis we
see that $\det(\vn_i\cdot\vn_j)=\det(n_i^a)^2=|(\vec{n}_1 \wedge
\vec{n}_2)\cdot\vec{n}_3|^2$. Thus we have relation (\ref{sqrt}).
Overall this leads to the measure \be
 d\mu={2 \over \pi^4}
{\prod_{I<J}\sin\t_{IJ}d\theta_{IJ} \over
{\sqrt{\det[\cos\t_{IJ}]}}}. \ee

\noindent\underline{Domain}: The domain of integration $\cD_\pi$
arise naturally since, for instance, the relation (\ref{rangle})
imply that \be \cos(\t_i+\t_j)\leq \cos\t_{ij} \leq
\cos(\t_i-\t_j), \ee which is equivalent to the condition \be
|\t_i-\t_j|\leq \t_{ij} \leq \pi -|\pi- \t_i-\t_j|. \ee

\noindent\underline{Integrand}: Finally, we get the theorem using
the character formula \be \chi_l(g(\t))= {\sin(l+1)\t \over
\sin\t}. \ee This completes the proof of the theorem.}

\noindent\textbf{Remarks:}

\noindent $1)$ It is interesting to note, by a direct computation,
that in the gauge where $g_0=1$ we have \be
\sqrt{\det[\cos\t_{IJ}]}={1\over 4} |\tr([g_1,g_2]g_3)| . \ee
Also, from relation (\ref{sqrt}) it is clear that if
$\det[\cos\t_{IJ}]=0$ then $\t_i=0,\pi$ or $|(\vec{n}_1 \wedge
\vec{n}_2)\cdot\vec{n}_3|=0$ in which case the tetrahedra is
degenerate. The reverse proposal is trivial.

\vspace{1ex} \noindent $2)$ The integration domain is invariant
under the transformation \be\label{sym}
\t_{IJ}\to\s_I\s_J\t_{IJ}+\frac{1-\s_I\s_J}{2}\pi, \ee with
$\sigma_I=\pm 1$. Under this transformation the Graham matrix
becomes $\sigma_I \sigma_J\cos\t_{IJ}$, and its determinant is
invariant, while the product of sinus produces a factor \be
\prod_{I<J}(\sigma_I\sigma_J)^{l_{IJ}}= \prod_I
\sigma_I^{\sum_{J\neq I}l_{IJ}}. \ee The exponent is always a even
integer due to the admissibility condition that the label around a
vertex must sum into an even integer. Using this extra discrete
symmetry we can restrict furthermore the integration domain to the
domain $\tcD_\pi$ which is the subset of $\cD_{\pi}$ such that
$\t_{i}\leq \pi/2$.

\vspace{1ex} \noindent $3)$ In this remark we clarify the fact
that we will consider the variables $\t_{IJ}$ both as lengths and
dihedral angles of spherical tetrahedra. There is a duality
between points and 2-spheres in $S^3$, namely if $g\in SU(2)$ we
can define the dual 2-sphere $S_g =\{x\in \SU (2), \tr(xg)=0\}$,
$S_g$ is the sphere of radius $\pi/2$ centered at $g$.
Reciprocally, such a sphere determines 2 points; its center $g$
and  its antipodal center $-g$. So, given a set of angles
$\t_{IJ}\in \cD_\pi$ we can construct -modulo gauge invariance-
$4$ group elements $g_I \in \SU(2)$ such that
$\cos\t_{IJ}=(1/2)\tr(g_{I}g_{J}^{-1})$. We can consider this
group elements as being the vertices of a spherical tetrahedron,
denoted $T(\t)$, so that $\t_{IJ}$ is the length of the edge
$(IJ)$. We can also consider these group elements as defining 4
spheres forming the 4 faces of a dual tetrahedron denoted
$T^*(\t)$. In this case $\t_{IJ}$ is the dihedral angle of the
edge dual to $(IJ)$ of the tetrahedron $T^*(\t)$. But given a
tetrahedron $T(\t)$ we can also compute the dihedral angles
$\ttt_{IJ}$ of this tetrahedron as in \ref{ttt}, and construct the
tetrahedron $T(\ttt_{IJ})$ : it turns out that we do not get a new
tetrahedron since $T(\ttt_{IJ})=T^*(\t_{IJ})$. It can be checked
that the Graham matrix of the dual tetrahedron
$T^*(\t_{IJ})=T(\ttt_{IJ})$ defined by $\ttG_{IJ}=\cos\ttt_{IJ}$
is related as follows to the Graham matrix  $G_{IJ}=\cos\t_{IJ}$.
Let $\t_{IJ}$ be in the interior of $\cD_\pi$ then $G$ is
invertible, $(G^{-1})_{II}>0$ and \be \ttG_{IJ} = {(G^{-1})_{IJ}
\over \sqrt{(G^{-1})_{II}(G^{-1})_{JJ}} }. \ee Finally if
$\t_{IJ}$ is in the boundary  of $\cD_\pi$, $T(\t)$ becomes a
degenerate spherical tetrahedron and $T^*(\t)$ is a flat
tetrahedron.


\subsection{Analysis of the integral}
Our aim is now to study the large N asymptotic of the integral
expressing the $6j$-symbol in terms of invariant variables
\begin{equation}
I(Nl_{IJ}) = {2\over \pi^4} \int_{\cD_\pi}  \frac{\prod_{I<J}
\sin((Nl_{IJ}+1)\t_{IJ})}{\sqrt{\det[\cos\t_{IJ}]}}
\prod_{I<J}d\t_{IJ}
\end{equation}
The form of the integrand -- an oscillatory function with an
argument proportional to $N$ times a function independent of $N$--
suggest that the asymptotic behavior can be studied by stationary
phase methods. However, this is not so simple since the integrand
is singular on the boundary of the integration domain. We
therefore have to make the asymptotic analysis by carefully taking
into account this singular behavior. For that purpose, we will
split the integration domain into two different parts, which will
be analyzed by adapted methods. We split the integration domain
$\cD_\pi$ into
\begin{equation}
\cD_\pi=\cD_{\pi,\e}^< \cup {\cD}_{\pi,\e}^> \ \ \ \mbox{with}\ \
\ \
\begin{array}{l}
\cD_{\pi,\e}^> = [\e,\pi-\e]^6 \cap \cD_\pi \\
{\cD}_{\pi,\e}^<=\cD_{\pi}-{\cD}_{\pi,\e}^>
\end{array}
\end{equation}
with $\epsilon>0$ is sufficiently small. Thus the integral
$I(l_{IJ})$ splits into $I^>(l_{IJ})$ and $I^<(l_{IJ})$
respectively given by the integrations on $\cD_{\pi,\e}^>$ and
${\cD}_{\pi,\e}^<$. Our main theorem summarizing the results
obtained for these asymptotic analysis is the following
\begin{theo}\label{maintheo}
Let $FT^*(l_{IJ})$ be the flat tetrahedron which is such that the
length of the edge $(IJ)$ is given by $l_{KL}$ (with $I,J,K,L$ all
distinct). Let $\T_{KL}$ be the angles between outward normals at
edge $(IJ)$ , and denote by $V^*(l_{IJ})$ its volume. Suppose that
$V^*(l)\neq 0 $ and that
$0<\epsilon<\mathrm{min}(\Theta_{IJ},\pi-\Theta_{IJ})$. Then the
first terms in the asymptotic expansion of both integrals are
\begin{equation}
I^{>}(Nl_{IJ}) \sim  -\frac{\sin(\sum_{I<J}(Nl_{IJ}+1)\T_{IJ})}{3
\pi N^3 V^*(l_{IJ})},\label{I>}
\end{equation}
and
\begin{equation}\label{I<}
I^{<}(Nl_{IJ}) \sim  \frac{1}{3 \pi N^3 V^*(l_{IJ})}.
\end{equation}
Overall leads to the asymptotic for the $6j$-symbol
\begin{equation}
 I(Nl_{IJ})\sim \frac{2}{N^3 3\pi  V^*(l_{IJ})} \cos^2\left(\sum_{I<J} \frac{(Nl_{IJ}+1)}{2} \T_{IJ}+{\pi \over 4}
\right).
\end{equation}
\end{theo}
This result is, of course, consistent with the results of
Ponzano-Regge and Roberts. The methods are however very different,
and to our taste simpler in the sense that they can be extrapolate
to the Lorentzian case as we will see in section
\ref{sec:lorentz}.

\subsection{Asymptotic of $I^>(Nl_{IJ})$}\label{section:I>}

\noindent\textbf{Method :} The result (\ref{I>}) is proved by the
stationary phase method. We recall first the main steps of this
method on a one dimensional example. First we will write the
integral under a form which is appropriate for the stationary
phase, namely something like
\begin{equation}
I_N=\int_a^b dx\ \phi(x) e^{iNf(x)},
\end{equation}
with $f$ an analytic function. This is done by proposition
\ref{prop:expexp}. The asymptotic expansion of such an integral is
given by contributions around the stationary points of the phase,
i.e. points in the complex plane such that $f'(x_0)=0$. In our
case such points are given by proposition \ref{prop:statpoints}.
The contribution of such a point is then obtained by expanding the
phase around it at second order and extend the integration to
infinity
\begin{equation}
I_N\sim\sum_{x_0} \int_{-\infty}^{+\infty} d(\d x)\ \phi(x_0)
e^{iN(f(x_0)+\frac{1}{2}f''(x_0)(\d x)^2)}
\end{equation}
$\phi(x)$ is the part of the integrand independant of $N$ and is
evaluated on the stationnary points $x_0$. The constant term can
be factored out of the integral and the integration is now a
gaussian which can be performed
\begin{equation}
I_N\sim \frac{1}{\sqrt{N}}\sum_{x_0} \phi(x_0)
e^{\frac{i\pi}{4}\mathrm{sgn}(f''(x_0))}
\sqrt{\frac{2\pi}{|f''(x_0)|}}e^{iNf(x_0)}
\end{equation}
Note that if $x_0$ lies on the boundary of the domain, its
contribution has to be half counted. For the n-dimensional
gaussian integration (see eq. \ref{eqn:gaussn}), we need to
compute the determinant and the signature of the quadratic form
obtained by the expansion. All of this is done in proposition
\ref{prop:expphase}. Recall that in the $n$ dimensional case, the
gaussian integration is performed using the fact that for a real
$n\times n$ symmetric invertible matrix $A$ with signature
$\sigma(A)$ we have
\begin{equation}
\int_{\real^n} [dX_i]\  \exp\left[i\left(\sum_{i,j} X_i A_{ij} X_j
\right)\right] = e^{i\sigma(A) \frac{\pi}{4}}
\sqrt{\frac{\pi^n}{|\det A|}}\label{eqn:gaussn}.
\end{equation}

\begin{prop}\label{prop:expexp}
The integral $I^>(Nl_{IJ})$ can be written as
\begin{equation}
I^{>}(Nl_{IJ}) = \frac{2 N^2}{(2\pi)^6} \sum_{\{\e_{IJ}=\pm\}}
\left(\prod_{I<J} \epsilon_{IJ}\right) I_{\e} \label{eqn:expexp}
\end{equation}
with
\begin{equation}\label{eqn:defIe}
I_{\e} = \int_{\real^4} \int_{\cD^>_{\e,\pi}} e^{i\sum_{I<J}
\e_{IJ}\t_{IJ}}\ e^{iN\left( \sum_{I<J} \e_{IJ} l_{IJ} \t_{IJ} +
\sum_{I,J} X_I \cos\t_{IJ} X_J \right)}d\t\ dX
\end{equation}
\end{prop}

\proof{We first split the sinuses into exponentials, which leads
to integrals which are still convergent. This is shown by notice
that due to the computation of the measure, the integral
\begin{equation}
\int_{{\cD}_{\pi,\e}^>} [d\t_{IJ}] \
\frac{1}{\sqrt{\det[\cos\t_{IJ}]}}
\end{equation}
is just the integral
\begin{equation}
\int_{{\cD}_{\pi,\e}^>} \frac{dh_1 dh_2 dh_3}{(\prod_i
\sin\t_i)(\prod_{i<j} \sin\t_{ij})}
\end{equation}
As the domain is a compact one, and exclude the points
$\t_i,\t_{ij}=0,\pi$, this integral is convergent. Now we use the
equation (\ref{eqn:gaussn}), to rewrite the denominator as arising
from such an integration (the signature of the matrix being 4).
One finally rescales the $X_I$ by $\sqrt{N}$.}

We consider now each of the integrals $I_{\e}$ and apply the
stationary phase method to compute their dominant asymptotic
contribution. The following proposition gives the stationary
points of the phase for these integrals. They are expressed in
terms of the geometrical elements of the tetrahedron
$FT^*(l_{IJ})$. In the following we denote $V^*$ its volume, $A_I$
its areas and $\T_{IJ}$ the angles between its outward normals.
\begin{prop}\label{prop:statpoints}
The stationarity equations of the integral $I_{\e}$ are given by
\begin{eqnarray}
\sum_J \cos\t_{IJ} X_J &=& 0, \label{c=0} \\
2 X_I X_J \sin\t_{IJ}&=& \e_{IJ} l_{IJ}. \label{s=0}
\end{eqnarray}
These equations admits solutions only when
$\e_{IJ}=(-1)^\n\s_I\s_J$ with $\s_I=\pm 1$ and $\n=0,1$. For each
of the 16 admissible choices of $\e_{IJ}$ there are 2 solutions
$s=\pm 1$ given by
\begin{eqnarray}
\t_{IJ}&=&\s_I\s_J\T_{IJ}+\frac{1-\s_I\s_J}{2}\pi, \\ X_I&=& s
i^{\n} \s_I\frac{A_I}{\sqrt{3V^*}}.
\end{eqnarray}
\end{prop}

\proof{ \underline{Localisation of the solutions in the complex
plane}: We begin by looking for general solutions in the complex
plane. Indeed in the method of stationary phase, one should look
for stationary points in the whole complex plane, as the contours
of integration can be analytically deformed in $\comp$. We prove
that a general solution $(\t_{IJ},X_I)$ of the stationarity
equations is in fact such that $\t_{IJ}\in\real$ and $X_I \in
\real$ or $i\real$. We start from the following lemma

\vspace{1ex}

\noindent{\textbf{Lemma:}} Consider $\a_{IJ}$ a $4 \times 4$
symmetric matrix such that $\sum_J \a_{IJ}=0$. Consider the
$5\times 5$ matrix $K(\a)$ defined by
\begin{equation}
K_{KL}=4\left(\a_{II}\a_{JJ}-(\a_{IJ})^2\right),\ \ \ K_{I5}=1,\ \
\ K_{55}=0,
\end{equation}
then we have
\begin{equation}\label{eqn:lemma}
(\mathrm{Cof}_{IJ})^2=8(\a_{IJ})^2\ \det K \label{eqn:transfoC}
\end{equation}
where $\mathrm{Cof}_{IJ}$ denote the $(IJ)$-cofactor of $K$.

\vspace{1ex}

We apply this lemma with $\a_{IJ}=X_I X_J \cos\t_{IJ}$ (the
hypothesis is satisfied due to the first stationary equation). In
that case $K$ is just the Cayley matrix $C_{IJ}=l^2_{KL}$ (due to
the second stationary equation), whose cofactors $\D_{IJ}$ are
real. As $\det C>0$, eq.\ref{eqn:transfoC} says that
$\a_{IJ}=X_IX_J\cos\t_{IJ}$ is real. $X_I X_J\sin\t_{IJ}$ is also
real from the second equation. Summing the square of these two
results,
\begin{equation}
0<(X_I X_J\sin\t_{IJ})^2+(X_IX_J\cos\t_{IJ})^2=(X_I X_J)^2
\end{equation}
we can conclude that $X_I X_J$ is real, and thus that $\t_{IJ}$ is
real. Lets denote $X_I=e^{\d_I} |X_I|$. If we look at the phase of
the second equation (\ref{s=0}) one gets that \be
e^{i\delta_I+i\delta_J}=\pm \epsilon_{IJ} \ee This imply that \be
e^{4i\delta_I}= \left( { \e_{IJ}\e_{IK}\over \e_{JK} }\right)^2=1.
\ee Therefore one has  $e^{i\delta_I}=i^\n \sigma_I$ with
$\sigma_I=\pm$, $\n=0,1$ and $ \epsilon_{IJ}=(-1)^\n
\sigma_I\sigma_J$. The conclusions of our study of the
localization of the solutions are that the only integrals with
stationary points are the 16 configurations such that $
\epsilon_{IJ}=(-1)^\n \sigma_I\sigma_J$ and that the solutions
$\t_{IJ}$ are real, while $X_I\in i^\n \sigma_I \real^+$.

\vspace{1ex}

\noindent\underline{Explicit solutions}: Now we compute the
explicit form of the solutions. Consider a solution for a
admissible configuration $\e_{IJ}=(-1)^\n\s_I\s_J$. The solutions
of this configuration and of the configuration all $\e_{IJ}=+1$
are mapped to each others by
\begin{eqnarray}
\t'_{IJ}&=&\s_I\s_J\t_{IJ}+\frac{(1-\s_I\s_J)}{2}\pi \nonumber \\
X'_I&=& i^\n \s_I X_I \label{eqn:mapsolutions}
\end{eqnarray}
Thus we only need to study the solution for the configuration
$\e_{IJ}=1$, the solutions for the others configuration will be
recovered by this map.

The matrix $G_{IJ}=\cos\t_{IJ}$ can be written as arising from the
scalar product of four unit vectors $\vn_I$ of $\real^4$, i-e
$G_{IJ}=\vn_I\cdot \vn_J$. The first relation (\ref{c=0}) imply
that $\det(G)=0$ so the $\vn_I$ span a vector space of dimension
at most three. It is easy to see that under the hypothesis of the
theorem it is exactly of dimension three. Lets denote $\vec{X}_I
=X_I \vn_I$, the second relation imply that $|\vec{X}_I \w
\vec{X}_J|=\frac{1}{2} l_{IJ}$. The LHS represent the area of the
triangle $(\vec{X}_I,\vec{X}_J)$. If the vector space was of
dimension two or less then these area, hence $l_{IJ}$, would be
related by triangular equalities. This is not allowed because the
tetrahedron $FT^*(l_{IJ})$ is supposed to be non-degenerate.
Therefore the matrix $G$ as exactly one null eigenvector.

Thus \cite{alek}, this matrix $G$ is the Graham matrix of a flat
non-degenerate tetrahedron, whose angles between outward normals
are the $\t_{IJ}$ and areas are proportional to the unique
eigenvector, the $X_I$ in our case. This tetrahedron is defined up
to scale. To fix the ideas, let us consider such a tetrahedron of
volume 1, let us call $\l_{IJ}$ its lengths, and $a_{I}$ its
areas, we thus have $X_I=\u a_I$. It is well known (see appendix
\ref{flat}) that these
geometrical elements satisfy \beq \sum_J \cos\t_{IJ} a_J =0, \\
a_I a_J \sin\t_{IJ}=(3/2)\l_{IJ}. \eeq By considering this
together with the second equation
\begin{equation}
l_{IJ}=2 \u^2 a_I a_J \sin\t_{IJ}=3 \u^2 \l_{IJ},
\end{equation}
we see that this tetrahedron is in fact homotethic to the
tetrahedron $FT^*(l_{IJ})$. The $\t_{IJ}$ solutions are thus
$\T_{IJ}$ the angles between outward normals of $FT^*(l_{IJ})$. As
its geometrical elements satisfy
\begin{equation}
A_I A_J\sin\T_{IJ}=\frac{3}{2} l_{IJ} V^*
\end{equation}
we have
\begin{equation}
X_I=s \frac{1}{\sqrt{3V^*}} A_I, \ \ \ s=\pm 1
\end{equation}
Finally for this configuration $\e_{IJ}=+1$, there are two
solutions. According to the map (\ref{eqn:mapsolutions}), for each
admissible configuration this gives the two solutions given in the
proposition .}

Our task is now to expand the phase around the solutions at the
second order, and compute the determinant and the signature of the
corresponding quadratic form.
\begin{prop}\label{prop:expphase}
For each of the 16 admissible configurations
$\e_{IJ}=(-1)^\n\s_I\s_J$ the expansion of the phase
\begin{equation}
\sum_{I<J} \e_{IJ} l_{IJ}\t_{IJ} + \sum_{I,J} X_I \cos\t_{IJ} X_J
\end{equation}
around any of the two $s=\pm 1$ solutions  gives
\begin{equation}
(-1)^\n\sum_{I<J} l_{IJ} \T_{IJ}+(-1)^\n \ Q(\d\t_{IJ},\d X_I)
\end{equation}
where $Q$ is a quadratic form. Its signature is $\sigma(Q)=2$ and
its determinant is $\det Q = (\frac{3}{2} V^*)^2$.
\end{prop}

\proof{ Consider first the expansion for $\e_{IJ}=+1$, the
constant term is easily seen to be $\sum_{I<J} l_{IJ}\T_{IJ}$. The
computation of expansion with fluctuations $(\d\t_{IJ},\d X_I)$ at
the second order leaves us with a quadratic form, which can be
expressed as such
\begin{eqnarray}
Q(\d\b_{IJ},\d
X_I)=&-&\sum_{I<J}\frac{1}{3V^*}A_IA_J\cos\T_{IJ}(\d\b_{IJ})^2 \nonumber\\
&+&\sum_{I} \left[2+\sum_{K\neq I}
(\frac{A_K}{A_I}\frac{1}{\cos\T_{IK}})\right](\d X_I)^2 \nonumber \\
&+&\sum_{I\neq J}\frac{1}{\cos\T_{IJ}}\ \d X_I\d X_J
\end{eqnarray}
where the $\d\b_{IJ}$ are a shift redefition of the $\d\t_{IJ}$

Consider now the expansion of the phase of an integral $I_\e$
around the corresponding solution. One has just to carefully add
the $\s_I$ and $\n$ into the previous results. The constant term
becomes
\begin{equation}
\sum_{I<J} \e_{IJ} l_{IJ}
(\s_I\s_J\t_{IJ}+\frac{(1-\s_I\s_J)}{2}\pi) =(-1)^\n \sum_{I<J}
l_{IJ} \t_{IJ} +\pi(-1)^\n\sum_{I<J} l_{IJ}\frac{\s_I\s_J-1}{2}
\end{equation}
The second term is always $0\ \mathrm{mod.}\ 2\pi$, due to the condition that the labels around
a face must sum into an even integer.
The quadratic form is unchanged, except one has just to redefine $\d X_I \to \s_I \d X_I$ to get
the same expression.

Our task is now to compute the determinant and the signature of
the quadratic form $Q$. This quadratic form is given by a diagonal
part (involving the $\d\b_{IJ}$), whose determinant is \be
{\left(\prod_{I<J} A_I A_J \cos\t_{IJ} \right) \over (3V^*)^6}.
\ee and a non-diagonal part which is the $4\times 4$ quadratic
form involving only the $\d X_I$. This non-diagonal part can by
computed in terms of the $A_I$ and $\cos\t_{IJ}$. Using the
formula (see the appendix \ref{flat}) \beq
A_I A_J \cos\t_{IJ}&=&-{1\over 16} \D_{IJ}\\
A_I^2&=&-{1\over 16} \D_{II}, \eeq  ($\D_{IJ}$ denotes the $(I,J)$
cofactor of the Cayley matrix), the total determinant can be
rewritten as
\begin{equation}
\det Q = \frac{1}{(3V^*)^6} P(\D_{IJ})
\end{equation}
with $P$ an homogeneous polynomial of order 6. The $\D_{IJ}$ can
now be expressed in terms of the $l_{IJ}$, the same can be done
for the volume $V^*$ and symbolic computation (Maple) allows to
check that
\begin{equation}
P(\Delta_{IJ})=\frac{(3V^*)^8}{4}
\end{equation}
Overall leads to
\begin{equation}
\det Q=\left(\frac{3}{2}V^*\right)^2
\end{equation}

The proposition concerning the signature of the quadratic form is verified only numerically.
We have computed numerically the signature of the quadratic form $Q$ for many randomly chosen
 tetrahedra. In all cases we obtain that $\sigma(Q)=2$.
}

We can now put together the results of the previous propositions
to prove the asymptotics of the integrals $I_{\e}$, then of the
total integral $I^>(Nl_{IJ})$. Consider an integral $I_{\e}$, if
the $\e_{IJ}$ can not be written $\e_{IJ}=(-1)^{\n} \s_I\s_J$, the
integral has no stationnary points and its asymptotics is a
$o(N^{-1/2})$. If $\e_{IJ}=(-1)^\n \s_I\s_J$, the oscillatory
phase has two stationnary points given by proposition
\ref{prop:statpoints}. Recall that the integrand of $I_\e$
contains a factor independant of $N$ (see eq.\ref{eqn:defIe})
which has to be evaluated on the stationnary points. This
evaluation is
\begin{equation}
exp\left[i\sum_{I<J} \e_{IJ}
\left(\s_I\s_J\T_{IJ}+(1-\s_I\s_J)\frac{\pi}{2}\right)\right]=e^{i(-1)^\n
\sum_{I<J} \T_{IJ}}\left(\prod_{I<J}\e_{IJ}\right)
\end{equation}
Recall that these points are the dihedral angles of a flat
tetrahedron and are on the boundary of the domain according to the
remark made after theorem 1. Their contributions have to be
half-counted. Using the results on the determinant and signature
of $Q$, we obtain that the asymptotics of $I_{\e}$ is
\begin{equation}
I_{\e}(Nl_{IJ})\sim
\frac{\pi^5}{N^5}\left(\prod_{I<J}\e_{IJ}\right)
\frac{2i(-1)^\n}{3V^*}e^{i(-1)^\n \sum_{I<J} (Nl_{IJ}+1)\T_{IJ}}
\end{equation}
Now there are 8 configurations of the $\e_{IJ}$ with $(-1)^\n=+1$
and 8 with $(-1)^\n=-1$. Summing this according to the expansion
(eq. \ref{eqn:expexp}), the total contribution is
\begin{equation}
I^>(Nl_{IJ}) \sim  -
\frac{1}{N^3}\frac{\sin(\sum_{I<J}(Nl_{IJ}+1)\T_{IJ})}{3\pi V^*}
\end{equation}
which proves the first part (\ref{I>}) of the theorem
\ref{maintheo}.


\subsection{Asymptotics of $I^<(Nl_{IJ})$} \label{sec:asI<}

\noindent\textbf{Method :} Before going on with the proof of the
second part (\ref{I<}) of the theorem let us illustrate the
problematic on a one dimensional example. Consider the simple case
of a one dimensional integral :
\begin{equation}
\int_a^b dx\ f(x) e^{iNx}
\end{equation}
If $f$ is a regular function on $[a;b]$, this integral has a
$O(\frac{1}{N})$ asymptotic which has its origin in the
contributions of the points $a,b$ of the lying on the boundary.
Now lets consider a function which is regular in $]a;b[$ but
posses a singularity on the boundary, for instance
\begin{equation}
\int_a^b dx\ \frac{f(x)}{(x-a)^{\a}} e^{iNx},
\end{equation}
with $f$ regular and non-vanishing at $a$, one can show  that the
asymptotic is an $O(1/N^{1-\a})$. This suggests that the leading
asymptotic behavior of an integral with singular points come from
the contributions around the most singular points. We expect that
in general the dominant contributions will come from singular
points of the integrant. It is easy to show that the zeros of
highest order of the denominator $\det[\cos\t_{IJ}]$, which will
dominate the large $N$ behavior of the integral $I^<(Nl_{IJ})$, is
given by the configuration where all $\t_{IJ}$ are zero. The proof
amounts to write the integral as a one dimensional integral which
realize an expansion around this singular configuration. We first
express the integral as a one-dimensional oscillatory integral
over an auxiliary variable $S$. This is done in proposition
\ref{prop:intS}. Then we compute the large $N$ asymptotics of this
integral in proposition \ref{prop:asymptS}. The result is thus
expressed as an integral over Euclidian tetrahedra, involving the
inverse of the volume $V$. This integral is exactly computed due
to a remarkable self-duality property given by theorem
\ref{th:sd}.

First it is useful to note that if $\e$ is small enough (i-e
$\e<\pi/3$) the domain $\cD^<_{\pi,\e}$ splits into eight
disconnected components which can be mapped into each other using
the discrete symmetry (eq. \ref{sym}). Using this property we have
\begin{equation}\label{eq:I<def}
I^<(Nl_{IJ})=\left({2\over \pi}\right)^4 \int_{\tcD_{\e}}
\frac{\prod_{I<J} \sin((
Nl_{IJ}+1)\t_{IJ})}{\sqrt{\det[\cos\t_{IJ}]}} \prod_{I<J}d\t_{IJ},
\end{equation}
 where
$\tcD_{\e}$ is the subset of $[0;\e]^6$ of $\t_{IJ}$ satisfying
the relations :
\be
\t_{IJ} \leq\t_{IK}+\t_{JK} \ee for any triple $(I,J,K)$ of
distinct elements. We can now rewrite the integral.

\begin{prop}\label{prop:intS}
\begin{equation}
I^<(Nl_{IJ})=\int_{-\infty}^{+\infty} e^{iNS} S^2 F_l(S) dS
\end{equation}
where $F_l(S)$ is a function of $S$ depending on the lengths
$l_{IJ}$ and given by \be\label{equation:Fl} F_l(S) =-{1\over 4
\pi^4}v.p \int_{\tcD_\e} \sum_{\e_{IJ}=\pm 1} \delta(S-
\sum_{IJ}\e_{IJ}l_{IJ}\t_{IJ}) {\prod_{I<J}
\e_{IJ}e^{i\e_{IJ}\t_{IJ}} \over S^{2} \sqrt{ [\det\cos \t_{IJ}]
}} \prod_{I<J} d\t_{IJ}, \ee where $v.p$ means that we take the
principal value of the integral i-e $v.p \int d\t
=lim_{\alpha\rightarrow 0} \int_{\t>\alpha} d\t$ and $\delta(x)$
is the Dirac functional.
\end{prop}
The rewriting of the integral is immediate, the key point here is
that the function $F_l(S)$ is well defined since the functionals
$S(\t_{IJ})=\sum_{IJ}\e_{IJ}l_{IJ}\t_{IJ}$ don not possess any
stationary points in the domain ${\tcD_\e}$  as was shown in the
section \ref{section:I>}.

The asymptotic expansion of the resulting one-dimensional integral
over $S$ can now be computed.

\begin{prop}\label{prop:asymptS}
\begin{equation}
I^<(Nl_{IJ})\sim \left({2\over \pi}\right)^3\frac{1}{N^3}
\int_{\tcD_{\infty}} du_{IJ}\ \frac{\prod_{I<J}
\sin(l_{IJ}u_{IJ})}{3\pi V(u_{IJ})}.\label{eq:deg}
\end{equation}
where $V(u_{IJ})$ is the volume of the flat tetrahedron which is
such that the length of the edge $(IJ)$ is given by $u_{IJ}$. The
domain $\tcD_{\infty}$ is the set of all  Euclidian tetrahedron.
\end{prop}

\proof{ It is well known that in the case
 of a one dimensional integral the asymptotic is:
\begin{equation}
I^<(Nl_{IJ}) =\int_{-\infty}^{+\infty}¥ dS e^{iNS}S^{2} F_l(S)
\sim \int_{-\infty}^{+\infty} dS e^{iNS}S^{2} F_l(0).
\end{equation}
Thus what we need is an evaluation at $S=0$ of the function
$F_l(S)$. This function indeed regular at this point, as it can be
seen by changing $\t_{IJ}=Su_{IJ}$. We get \be \label{eq:Flexpand}
F_l(S) = -{1\over 4 \pi^4} S^3 v.p \int_{\tcD_{\e/S}}
\sum_{\e_{IJ}=\pm 1} \delta(1- \sum_{IJ}\e_{IJ}l_{IJ}u_{IJ}) {
\prod_{I<J} \e_{IJ}e^{iS\e_{IJ}u_{IJ}}  \over
\sqrt{[\det\cos(Su_{IJ})]} }  \prod_{I<J} du_{IJ}. \ee As $S\to
0$, the domain becomes the domain of all Euclidian tetrahedra. The
denominator can now be expanded as $S\to 0$. The proposition
follows from this expansion
\begin{equation}
\sqrt{\det[\cos(Su_{IJ})]} \sim_{S\to 0} S^3 6 V(u_{IJ})
\end{equation}
where $V(u_{IJ})$ denotes the volume of tetrahedron, whose length
$(IJ)$ is $u_{IJ}$.
This lemma can be proved by direct expansion,
but if follows from this observation : Consider the determinant
$\D=\det\cos\t_{IJ}$ (with $\t_{0i}=\t_i$). By substracting row 4
(resp. line 4) to the three others, one obtains the determinant
\begin{equation}
\D=\det(\cos\t_{ij}-\cos\t_i\cos\t_j)
\end{equation}
which is rewritten, following relation (eq.
\ref{eqn:sphericaldual}),
\begin{equation}
\D=(\prod_i \sin^2\t_i) \det(\vn_i\cdot\vn_j)
\end{equation}
The limit $\t_i\to 0$ of this expression leads to the expression
$\det(\t_i \t_j (\vn_i\cdot\vn_j))$ which is the square of the
volume of the flat tetrahedron defined by vectors
$\t_1\vn_1,\t_2\vn_2,\t_3\vn_3$ (see eq.\ref{volumetetra}). Thus
we have the property
\begin{equation}
\det[\cos\t_{IJ}] \sim_{\t\to 0} 6^2 V^2
\end{equation}
with $V$ the volume of the tetrahedron whose lengths are
$(\t_i,\t_{ij})$.}

We have express the asymptotic of the integral $I^<(Nl_{IJ})$ as
an integral (\ref{eq:deg}). This integral can be in fact exactly
computed, showing a self-duality property of the inverse of the
volume of the tetrahedron.

\begin{theo}\label{th:sd}
Let $V(u_{IJ})$ (resp. $V^*(l_{IJ})$) be the volume of the flat tetrahedron which is such that the length of the edge
$(IJ)$ is given by $u_{IJ}$ (resp. $l_{KL}$)
The inverse volume satisfy the remarkable self duality property
\begin{equation}\label{sd}
\left({2\over \pi}\right)^3\int_{\tcD_{\infty}} du_{IJ}\
\frac{\prod_{I<J}\sin(l_{IJ}u_{IJ})}{3\pi V(u_{IJ})} =
\frac{1}{3\pi V^*(l _{IJ})}.
\end{equation}
\end{theo}

The proof of this theorem make the use of the following
proposition concerning the measures on the spaces of Euclidian
tetrahedra.
\begin{prop}\label{vu}
Let $\vu_i$ be 3 vectors of $\real^3$, and lets denote $u_{0i}=|\vu_i|$, $u_{ij}=|\vu_i-\vu_j|$
then
\begin{equation}
\prod_i {d^3\vu_i\over 2\pi} = \frac{ \prod_{I<J} u_{IJ}\ du_{IJ}}{3 \pi V(u_{IJ})}
\end{equation}
where  $V(u_{IJ})$ is the volume of the
tetrahedron $(\vu_1,\vu_2,\vu_3)$ whose edges length are $u_{IJ}$.
\end{prop}
\proof{ The measure on $\real^3$ is given by $d^3\vec{u}_i= 4\pi
u_i^2 du_i d^2\vec{n}_i$, where $u_i = |\vu_i|$, $\vec{n}_i\in
S^2$ and $d^2\vec{n}$ is the normalized measure on the 2-sphere.
So the measure $d\mu= \prod_i \frac{d^3\vu_i}{2\pi} $ becomes \be
d\mu = 2^3 \left(\prod_{i=1}^3 u_i^2 du_i\right) d^2\vn_i. \ee
From there we proceed exactly as in the proof of theorem 1, using
the fact that \be 6 V(u_{IJ})=(\prod_i u_i)
\det(\vn_1\wedge\vn_2\cdot\vn_3).\ee }

\noindent We can now prove the theorem. This proposition implies
that the integral eq. \ref{eq:deg} has  a form similar to eq.
\ref{6j} \be I^<(l_{IJ})\sim\left({2\over \pi}\right)^3
\frac{1}{(2\pi)^3} \int_{\real^3} \prod_i {d^3\vu_i} \prod_i
\frac{\sin(l_i|\vec{u}_i|)}{|\vec{u}_i|} \prod_{i<j}
\frac{\sin(l_{ij}|\vec{u}_i-\vec{u}_j|)}{|\vec{u}_i-\vec{u}_j|},
\ee where $G$ is replaced by $\real^3$ and the character
$\chi_l(g)$ is replaced by $K_l(\vec{u})=\sin(l|\vec{u}|)/
|\vec{u}|$. This fact was already conjectured in the paper of Baez
et al \cite{Baez1} where the kernel $K_l(\vec{u})$ has been
interpreted in term of spin networks of the Euclidean group.
Changing the lengths to their dual $l_i=L_{jk},l_{ij}=L_{k}$ and
using the Kirillov formula
\begin{equation}
\frac{\sin L|x|}{|x|}= L \int_{S^2} e^{iL
\vec{x}\cdot\vec{n}}d^2\vec{n}
\end{equation}
one obtains \be I^<(l_{IJ})= \frac{1}{\pi^6} (\prod_{I<J} L_{IJ})
\int_{\real^3} \prod d^3\vec{u_i} \int_{(S^2)^6} (\prod_{I<J}
d^2n_{IJ})
\prod_k\exp\left\{i\vec{u}_k\cdot\left(L_{ij}\vec{n}_{ij}+L_{j}\vec{n}_{j}-L_{i}\vec{n}_{i}\right)
\right\} \ee The integrals over the sphere can be rewritten as
integrals over $\real^{3}$:
\begin{equation}
\int_{S^2} d^2\vec{n}=\frac{1}{4\pi L^2}\int_{\real^3} d^3\vec{X}\
\d(|\vec{X}|-L),
\end{equation}
while the integrals over the $\vec{u}_i$ can be performed $\int
e^{i\vec{u}\vec{x}} d^{3}\vec{u}=(2\pi)^{3} \delta^{3}(\vec{x})$,
leading to others $\d$-functions.
\begin{equation}
\frac{1}{\pi^6} \frac{(2\pi)^9}{(4\pi)^6} \int_{(\real^3)^6}
\prod_i d^3X_i \prod_{i<j} d^3X_{ij} \prod_{i<j}
\d^{(3)}\left(\vec{X}_{ij}-(\vec{X}_i-\vec{X}_j)\right) \prod_i
\frac{1}{L_i} \d(|\vec{X}_i|-L_i) \prod_{i<j} \frac{1}{L_{ij}}
\d(|\vec{X}_{ij}|-L_{ij})
\end{equation}
The integrations on $\vec{X}_{ij}$ can be performed, due to the $\d^{(3)}$ functions.
It remains
\begin{equation}
I^<(l_{IJ})=\frac{1}{(2\pi)^3} \int_{(\real^3)^3} {d\vec{X}_i}
\prod_i \frac{1}{L_i} \d(|\vec{X}_i|-L_i) \prod_{i<j}
\frac{1}{L_{ij}} \d(|\vec{X}_{i}-\vec{X}_j|-L_{ij})
\end{equation}
Changing into variables
$X_i=|\vec{X}_i|,X_{ij}=|\vec{X}_i-\vec{X}_j|$ and using again the
proposition \ref{vu} one gets finally \be I^<(Nl_{IJ})\sim
\frac{1}{3 \pi N^3 V(L_i,L_{ij})} = \frac{1}{3 \pi N^3
V^*(l_{IJ})} \ee

\noindent\textbf{Remark :} The self duality property (\ref{sd}) of
the inverse volume shares strong similarity with the self duality
property of the square of the quantum $6j$ symbol which has been
shown by Barrett \cite{Barrettsd}. It is therefore tempting to ask
whether this property has a natural interpretation in terms of
quantum groups where the square root of the inverse volume would
be interpreted as a quantum $6j$ symbol.


\section{Lorentzian 6j symbol}\label{sec:lorentz}
In \cite{Flor} it was shown that the partition function and
transition amplitudes of Lorentzian 3d gravity can be written as a
state sum model. To construct this state sum model one
triangulates a 3d manifold, one gives an orientation to the edges
of the triangulation and color these edges by unitary
representation of $\SL(2,\real)$. The weight associated with an
oriented colored tetrahedron is the corresponding  $6j$ symbol of
$\SL(2,\real)$. Among all the possible representations of
$\SL(2,\real)$ only the positive discrete series, the negative
discrete series and the principal series play a role. All these
representations are unitary and infinite dimensional \cite{Knapp}.
We will denote $T_{il^+}$, $l \in \Nat$, the positive discrete
series of weight $l$, $T_{-il^-}$, $l \in \Nat$, the negative
discrete series of weight $l$, and $T_{\rho}$, $\rho \in \real^+$,
the principal series of weight $\rho$. To simplify the exposition
we have adopted a
 unifying notation, all the representations are denoted $T_\l$ where
 $\l =+il^+, -il^-$ or $\rho$ refers respectively to
 positive discrete series, negative discrete series and principal series.
Similarly to the case of $\SU(2)$ one can construct the
Clebsch-Gordan coefficients, and $6j$-symbols as a recombination
of Clebsch-Gordan coefficients (we refer the reader to
\cite{Davids} for a precise definition of these objects). Due to
the fact that we can color each oriented edge of the tetrahedron
by either $il^+,-il^-$ or $\rho$ there are many different types of
6j symbols. However the square of all these 6j symbols can be
expressed by an integral formula similar to eq.\ref{6j}
\begin{equation}\label{6jLor}
I(\l_{IJ})= \left|\left\{\begin{array}{ccc}
\l_{01} & \l_{02} & \l_{03}\\
\l_{23}& \l_{13} & \l _{12}
\end{array}\right\}\right|^2
= \int_{(\SL(2,\real))^4}\ \prod_{I<J} \chi_{\l_{IJ}}(g_J g_I^{-1}) [\prod_I dg_I]
\end{equation}
where $\chi_{\l}(g)$ is the character  corresponding to the $\l$
representation (see appendix \ref{sl}). In the following we will
suppose that the tetrahedra is ordered by the vertices order, i-e
$(IJ)$ is positively oriented if $I<J$. The triples $\l_{IJ},
\l_{IK}, \l_{IL}$ are supposed to satisfy some admissibility
conditions described in the appendix \ref{sl}. The measure
$[\prod_I dg_I]$ denotes the, fully gauge fixed, product of Haar
measures. It is here necessary to fully gauge fix the symmetry $
g_I\to k g_I h$ in order to avoid divergences since the group is
non compact. Using techniques similar to the one we used in the
previous section we can compute the gauge fixed measure
\begin{prop}\label{lormeas}
\be
[\prod_I dg_I]=
 \left(\prod_{I<J} d\tau_{IJ} \right)\frac{|\sinh(\tau_{IJ})|}
{\sqrt{|\det[G_{IJ}(\tau)]}|}
\ee\ where $\tau_{IJ}= i\theta_{IJ}$ and $G_{IJ}=\cos\t_{IJ}$
if $ g_Jg_I^{-1}$ is elliptic and conjugated to $k_{\t_{IJ}}$ in this case
 $d\tau_{IJ} =d\t_{IJ}$; and
$\tau_{IJ}= |t_{IJ}|$, $G_{IJ}=\v_{IJ} \cosh t_{IJ}$ if $
g_Jg_I^{-1}$ is hyperbolic and conjugated to $\v_{IJ} a_{t_{IJ}}$,
$\v_{IJ}=\pm 1$, see appendix \ref{sl} for the notations.
\end{prop}
These gauge fixed variables are interpreted as parametrizing the
invariant geometry of an AdS tetrahedron. The technics used to
deal with non compact spin network integrals where first
introduced in \cite{FL}. Similarly to the case of $\SU(2)$ one can
split the integral expression of the Lorentzian $6j$ into a sum
$I^>(\l_{IJ})+I^<(\l_{IJ})$ where $I^> $restrict the integration
range into a domain where $|\tau_{IJ}|>\e$ and $|i\pi
-\tau_{IJ}|>\e$. In the following we analyze the asymptotic
behavior of $I^>(N\l_{IJ})$ when $N\to \infty$. In order to
illustrate our method, we will treat in detail the case where all
$\lambda_{IJ}$ are real, corresponding to a spacelike Lorentzian
tetrahedron, and the case where all $\l_{IJ}$ are imaginary,
corresponding to timelike Lorentzian tetrahedron. The last
subsection gives a method to solve the stationnarity equations in
the general case.

\subsection{Spacelike tetrahedra}
The case where all representations are continuous is simpler
since, first there is no restriction on the set $\rho_{IJ}$ and
second  the characters $\chi_{\rho}$ are supported only on
hyperbolic elements. The integral reads \be I^>(\r_{IJ})=
\sum_{\nu_{IJ}=\pm} \int_{t_{IJ}>\e} \frac{\prod_{I<J}
\cos(\r_{IJ}t_{IJ})} {\sqrt{|\det[\nu_{IJ}\cosh
t_{IJ}]}|}[\prod_{I<J} dt_{IJ}] . \ee We have $\v_{II}=+1$ and
$\v_{IJ}$ are restricted to the ones actually arising from an AdS
tetrahedron. The cosine can be expanded and the denominator
expressed as a Gaussian integral, overall this leads to \be
I^>(N\r_{IJ})= N^2 \left({1\over 2}\right)^6 {1\over \pi^2}
\sum_{\e_{IJ} =\pm, \nu_{IJ}=\pm} \int_{\real^4} \int_{t_{IJ}>\e}
e^{iN S_{\e \r, \nu}(t_{IJ},X_I) } (\prod_{I<J} d t_{IJ})(\prod_I
dX_I), \ee where the oscillating phase is \be S_{\e_{IJ}
\r_{IJ},\nu_{IJ}}(\ t_{IJ},X_I) =\sum_{I<J} \e_{IJ} \r_{IJ} \
t_{IJ} + \sum_{I,J} X_I \nu_{IJ} \cosh t_{IJ} X_J. \ee  The
asymptotic behavior of this integral is driven by the stationary
points of this action. They satisfy \beq \sum_J \nu_{IJ} \cosh
t_{IJ} X_J &=& 0, \label{ch=0}
\\ 2 X_I X_J \sinh t_{IJ}&=& -\nu_{IJ} \e_{IJ} \rho_{IJ}. \eeq The
analysis of this system of equation is similar to the one
performed in proposition \ref{prop:statpoints}. This analysis will
be done in subsection \ref{generalL} in the general case of a
Lorentzian tetrahedron. The conclusions are given in terms of the
geometrical elements of the tetrahedron $\cT(\rho)$ given by the
spacelike lengths $\rho_{IJ}$. Let us denote $A_I$ its areas,
$V^*$ its volume and $(V_{IJ},T_{IJ})$ its Lorentzian dihedral
angles. The possible solutions are given by \beq X_I &=& \pm
i^\eta \sigma_I {A_I\over \sqrt{3V^*}}\\ t_{IJ}&=& T_{IJ}\eeq The
expansion of the phase around the stationnary point $T_{IJ}$ gives
the constant term \be S=\sum_{I<J}\e_{IJ} \rho_{IJ} T_{IJ} \ee The
quadratic form obtained by expansion around these configurations
has a determinant which equals to $- ({3\over 2}V^*)^2$. In order
to prove the result about the determinant one first write the
quadratic form in term of cofactors $ \D_{IJ}$ of the Cayley
matrix (see appendix \ref{flat}). The form we obtain is the same
as the one obtained in the $\SU(2)$ case except for a global minus
sign in front of the non-diagonal part. It was shown there that
the determinant of the quadratic form is proportional to the
fourth power of determinant of the Cayley matrix. Since this is a
polynomial identity it extend when the Cayley matrix comes from a
Lorentzian tetrahedron. Numerical simulations suggest that the
signature of this quadratic form is $\pm 2$ depending on the
number of positive and negative $V_{IJ}$ in the dihedral angles of
the spacelike tetrahedron $\cT(\rho)$. We now have to consider the
compatibility equations on the signs to actually determine the
possible $\s_I,\v_{IJ},\e_{IJ},\n$. The first stationnarity
equation can be written
\begin{equation}
\sum_J A_J \s_J \v_{IJ} \cosh T_{IJ} = \sum_J A_J \s_J \v_{IJ}
V_{IJ} G_{IJ}=0\label{eqn:compat1}
\end{equation}
where $G_{IJ}=V_{IJ}\cosh T_{IJ}$ is the Graham matrix of the
tetrahedron $\cT(\rho)$, while the sign of the second equation
gives
\begin{equation}
\e_{IJ}=-(-1)^\n \v_{IJ}\s_I \s_J\label{eqn:compat2}
\end{equation}
Considering the equation (\ref{eqn:compat1}), we know (see
appendix \ref{flat}) that the $V_{IJ}$ can be written
$V_{IJ}=-\a_I\a_J$. If we consider the factorizable configurations
of $\v_{IJ}=\b_I \b_J$, we get for the equation
(\ref{eqn:compat1})
\begin{equation}
-\a_I\b_I \sum_J A_J \s_J \a_J \b_J G_{IJ}=0
\end{equation}
The fact that $G_{IJ}$ possess only one null eigenvector given by
the $A_J$ amounts to identify $\b_J=\s_J \a_J$ and the
contributing $\v_{IJ}$ as $\v_{IJ}=-\s_I\s_J V_{IJ}$. Considering
now the second compatibility equation (\ref{eqn:compat2}) gives
the possible $\e_{IJ}$ as $\e_{IJ}=(-1)^\n V_{IJ}$. Taking into
account theses contributions, we obtain the following asymptotic
term \be \pm {\cC \over N^3} \frac{\sin(\sum_{I<J} V_{IJ}
\rho_{IJ} T_{IJ})}{3\pi V^*} \ee for $\cC$ a constant independant
of the $\rho_{IJ}$.

\subsection{Timelike tetrahedra}
In the case where the representations are discrete and equal to
$l_{IJ}^{\e_{IJ}}$, the integral reads \be\label{int:lor}
I^>(iN\e_{IJ} l_{IJ}^{\e_{IJ}})= \int_{\cD_\e} \left(\prod_{I<J}
{-\e_{IJ}  \over 2i}{|\sin\t_{IJ}|\over \sin\t_{IJ}}\right)
 \frac{ e^{i \sum_{IJ}
\e_{IJ}(Nl^{\e_{IJ}}_{IJ}-1)\t_{IJ}}}{\sqrt{|\det[\cos \t_{IJ}]}|}
[\prod_{I<J} d\t_{IJ}] + \cdots \ee where the domain of
integration is the set of angles $ \t_{IJ}\in [\e,\pi-\e] \cup
[-\pi +\e,-\e]$ which are such that $\det[\cos \t_{IJ}]$ is
negative. It corresponds to the set of timelike AdS tetrahedra.
The dots refer to the fact there are other terms in the integral
coming from the sectors  were at least one element $g_Jg_I^{-1}$
is hyperbolic. These terms contain, in the integrand, at least one
factor $ \exp{-N|t_{IJ}|}$, such a factor is bounded by
$\exp{-N\e}$ and is exponentially small, therefore they  do not
contribute to the asymptotic behavior of the integral. Even if the
expression eq.\ref{int:lor} is suitable for the asymptotic
analysis, it is interesting to note that we could restrict the
integration to be the set of angles for which $ \t_{IJ}\in
[\e,\pi-\e], \det[\cos \t_{IJ}]<0$ if we use the identity
 \be
 \int_{-\pi}^{+\pi}P(\t) {|\sin\t|\over \sin\t} {e^{iK\t} \over 2i}\ d\t =
 \int_0^\pi P(\t) \sin(K\t)\ d\t.
 \ee
for $P$ such that $P(-\t)=P(\t)$, which is the case of the
denominator for each of its arguments $\t_{IJ}$. The integral now
reads \be I^>(iN\e_{IJ} l_{IJ}^{\e_{IJ}})= \int_{\cD_\e}
\prod_{I<J} {\sin((Nl^{\e_{IJ}}_{IJ}-1) \t_{IJ})  \over
\sqrt{|\det[\cos \t_{IJ}]}|},
 \ee
a form which is strikingly similar to the Euclidean integral
eq.\ref{int:eucl}.

The analysis of the asymptotic behavior goes along the same line
as before and the stationary points are in one to one
correspondence with timelike tetrahedra. Modulo permutation of the
vertices and change of orientation one can always present such a
tetrahedron in a form where all the edges are future timelike
vectors and the vertex 3 is in the future of 2 which is in the
future of 2... in the future of 0. This amount to take al
$\l_{IJ}, I>J$ to label positive discrete representation. For such
a configuration and using again the same technics one expect the
asymptotic to be \be I^>(iNl_{IJ}^+) \sim \frac{\cC}{N^3} {e^{i
\sum_{I<J}(Nl^+_{IJ}-1) \T_{IJ}}\over V^*(l)} + c.c \ee where
$\T_{IJ}$ are the dihedral angles of the timelike tetrahedron
whose edge are $l_{IJ}^+$, $V^*$ is its volume. $\cC$ is a
constant independent of the l$_{IJ}$ and c.c stands for complex
conjugate.

\subsection{General method for the stationarity
equations}\label{generalL}

In the general case of Lorentzian 6j symbol, the edges are
labelled by (discrete or continuous) representations. If the
representation is continuous then the character is zero on
elliptic elements and if the representation is discrete the
character is exponentially small on hyperbolic elements. Thus the
only case of interest for the asymptotic behavior is when the
integration is over elliptic elements if the edge is labelled by a
discrete representation and over hyperbolic elements if the edge
is labelled by a continuous representation. This amounts to
integrate over AdS tetrahedra which have spacelike edge when the
representation is continuous and timelike edge when the
representation is discrete. The asymptotic behavior of this
integral is controlled by the solutions of the following
stationary equations.
\begin{eqnarray}\label{eq:lorstat}
\sum_J X_J \v_{IJ} \cosh\tau_{IJ} &=& 0 \label{eq:ch}\\ 2  X_I X_J
\v_{IJ} \sinh\tau_{IJ} &=& -\e_{IJ}\l_{IJ},\label{eq:sh}
\end{eqnarray}
In the elliptic case, $\l_{IJ}$ stands for $\pm i l_{IJ}$,
$\tau_{IJ}$ denotes $i\t_{IJ}$ and $\v_{IJ}$ is always $1$. In the
hyperbolic case $\l_{IJ}=\rho_{IJ}$, $\tau_{IJ}=t_{IJ}$ and
$\v_{IJ}=\pm 1$ for the different sectors of hyperbolic angles. We
will express the solutions of these equations in terms of the
geometrical elements of the Lorentzian tetrahedra $\cT(\l)$ given
by the square lengths $\l^2_{IJ}$. Let us denote by $\T_{IJ}$ and
$(V_{IJ},T_{IJ})$ its dihedral angles respectively in the timelike
and spacelike case. Let us denote $A_I$ its areas and $V$ its
volume. We will procede as in the $\SU(2)$ case by first
identifying where the solutions lie in the complex plane, then
compute them explicitely.

\noindent\underline{Localization:}  We can repeat the analysis
done in the $\SU(2)$ case. We apply the same lemma
(eq.\ref{eqn:lemma}) to obtain
\begin{equation}\label{XXc}
(X_I X_J \cosh \tau_{IJ})^2=\frac{(\D_{IJ})^2}{-8 \det C(\l)}
\end{equation}
with $C(\l)$ the Cayley matrix of the $\l_{IJ}$ and $\D_{IJ}$ its
cofactor. This cofactor is real and the determinant of $C$ is
negative. We can conclude that $X_I X_J \cosh\tau_{IJ}$ is real.
Now the second stationnarity equation tells us that
$X_IX_J\sinh\tau_{IJ}$ is real in the spacelike case and pure
imaginary in the timelike case. Taking the difference of the
square of these two equations gives
\begin{equation}
(X_IX_J)^2=(X_I X_J \cosh \tau_{IJ})^2 - (X_IX_J\sinh\tau_{IJ})^2
= \frac{(\D_{IJ})^2}{-8 \det C(\l)} - \frac{\l_{IJ}^2}{4}
\end{equation}
One can conclude that $X_IX_J$ is real since the RHS of this
equation is always positive. This is clear in the timelike case
since the square of $\l_{IJ}=il_{IJ}$ is negative. In the
spacelike case, this arise from rewriting it using the relations
(\ref{wedge}) and (\ref{co})
\begin{equation}
\frac{(\D_{IJ})^2}{-8 \det C(\l)} - \frac{\l_{IJ}^2}{4} =
\frac{A_I^2 A_J^2}{(3V)^2} \cosh^2 T_{IJ}-\frac{A_I^2
A_J^2}{(3V)^2} \sinh^2 T_{IJ}= \frac{A_I^2 A_J^2}{(3V)^2}
\end{equation}

We thus know that $X_I X_J$ is real, $\cosh\tau_{IJ}$ is real,
$\sinh\tau_{IJ}$ is real in the spacelike case and pure imaginary
in the timelike case. This allows to conclude that $\tau_{IJ}$ is
pure imaginary in the timelike case (hence $\t_{IJ}$ is real) and
$\tau_{IJ}$ is real modulo $i\pi$ in the spacelike case. However
this $i\pi$ ambiguity corresponds to a $\pm$ ambiguity in the sign
of $\cosh\tau_{IJ}$ and $\sinh\tau_{IJ}$, and thus leads to the
same solution obtained by taking $\tau_{IJ}$ and changing the sign
of $\v_{IJ}$. These solutions are already taken into account and
we can restrict to the case of $\tau_{IJ}$ real in the spacelike
case. This completes our investigation of the localization of the
solutions.

\noindent\underline{Explicit solutions:} We consider first the
equation (\ref{XXc}) for $I=J$ and use the equations (\ref{co})
and (\ref{V}) to rewrite the RHS. We obtain
\begin{equation}
X_I^4=\frac{A_I^4}{(3V)^2}
\end{equation}
Thus it exists $\s_I=\pm 1$ and $\n_I=0,1$ such that $X_I=\s_I
i^{\n_I}/\sqrt{3V}$. The fact that $X_IX_J$ is real for all $IJ$
leads to the fact that all $\n_I$ are equal and
\begin{equation}
X_I=\s_I i^\n \frac{A_I}{\sqrt{3V}}
\end{equation}

Now we consider the equation (\ref{XXc}) for $I\neq J$ and use the
solution we found for the $X_I$ to get
\begin{eqnarray}
(\cos\t_{IJ})^2&=&(\cos\T_{IJ})^2 \\
(\cosh t_{IJ})^2 &=& (\cosh T_{IJ})^2
\end{eqnarray}
We have to solve these equations in $\real$, but since the
original integral is symmetric by changing $t \to -t$ and $\t \to
\t + 2\pi$, we keep only the actually different solutions and we
solve these equations for $t\in \real^+$ and $\t\in [-\pi;\pi]$.
This gives the solutions
\begin{eqnarray}
t_{IJ} &=& T_{IJ} \\
\t_{IJ} &=&
\a_{IJ}\left(\b_{IJ}\T_{IJ}+(1-\b_{IJ})\frac{\pi}{2}\right) \ \
\mbox{ for } \a_{IJ},\b_{IJ}=\pm 1
\end{eqnarray}
Reporting these result in the first stationnaity equation gives
\begin{equation}
\sum_J \s_J A_J \left\{\begin{array}{l} \v_{IJ}\cosh T_{IJ} \\
\b_{IJ}\cos\T_{IJ}
\end{array}\right\}=0
\end{equation}
We still have to determine the possible $\v_{IJ},\s_I,\b_{IJ}...$
by considering compatibility equations. In the previous one, one
can use the fact that by hypothesis, the Graham matrix of
$\cT(\l)$
\begin{equation}
G_{IJ}=\left\{\begin{array}{l} V_{IJ}\cosh T_{IJ} \\
\cos\T_{IJ}
\end{array}\right\}
\end{equation}
possess only one null eigenvector given by the $A_I$. On the other
hand, the sign of the second stationnary equation gives also
equations
\begin{equation}
(-1)^\n \s_I \s_J \left\{\begin{array}{l} \v_{IJ} \\ \a_{IJ}
\end{array}\right\}=-\e_{IJ}
\end{equation}
which should be satisfy in order for the solution to exists. The
bottom line is that the system of equation \ref{eq:lorstat} admits
solutions only if $\l_{IJ}$ could be interpreted as a set of
length of an oriented tetrahedra $\cT(\l_{IJ})$. In that case the
on-shell action for this solutions is, up to a sign, proportional
to the Regge action of Lorentzian tetrahedra. The determinant of
the quadratic form obtained by looking at fluctuation around these
configuration is proportional to the volume of the tetrahedron.


\section{10j symbol}\label{sec:10j}
\subsection{Integral expression for the 10j-symbol}
To analyze the case of the 10j-symbol along the same lines of the
6j-symbol, we will use higher-dimensional generalizations of
propositions used in the previous parts. The analog of the theorem
expressing the symbol as an integral over invariant variables is
\begin{theo}
The 10j-symbol can be expressed as the following integral \be
{4\over \pi^6}\int_{\cD'_{\pi}} \left(
\prod_{I<J}\sin((l_{IJ}+1)\t_{IJ})\right)  \delta(G(\t_{IJ}))\
[\prod_{I<J} d\t_{IJ}],
 \ee
 where $G(\t)$ denotes the determinant of the $5\times 5$ Graham matrix $G_{IJ}=\cos\t_{IJ}$
 associated with the spherical 4-simplex $01234$ and $\delta$ is the Dirac
 functional. $\cD'_{\pi}$ is the set of all spherical 4-simplices.
\end{theo}

\proof{
We start from the Barrett's expression for the 10j-symbol \cite{Barint}
\begin{equation}
\int_{\SU(2)^5}  \chi_{l_{IJ}}(g_Ig_J^{-1})[\prod_I dg_I]
\end{equation}
Then we apply the same method as in the 6j case. By gauge fixing
$g_0=1$ we are left with an integral over $\SU(2)^4$ with an
$AdG$-invariance. We have seen in the proof of theorem 1 that if
$g_1,g_2$ are two groups elements we can express their invariant
measure in term of the angles $\t_{1},\t_2,\t_{12}$. \be
dg_1dg_2={2\over \pi^2}\sin\t_1 d\t_1 \sin\t_2 d\t_2 \sin
\t_{12}d\t_{12} \ee Moreover if $ g_I$ is an additional group
element its measure is given by \be\label{eq:invmeas}
dg_I={1\over\pi^2} { \sin\t_I d\t_I \sin\t_{1I} d\t_{1I} \sin
\t_{2I}d\t_{2I} \over \sqrt{G_{12I}} } \ee where $G_{12I}$  is the
determinant of the Graham matrix for the spherical tetrahedron
whose vertices are $1,g_1,g_2,g_I$. using these results for
$g_3,g_4$ allows to prove the following  integral expression for
the 10j-symbol
\begin{equation}\label{eq:4s}
{2\over \pi^6}\int_{\cD'_{\pi}} [\prod_{\cI} d\t_{IJ}]\
\frac{\prod_{\cI} \sin(l_{IJ}+1)\t_{IJ}}{\sqrt{\L_{33}\ \L_{44}}}
\ \frac{\sin(l_{34}+1)\t_{34}}{\sin\t_{34}}
\end{equation}
where the set $\cI$ is the set $1\leq I < J \leq 4, (I,J)\neq
(3,4)$. $\L_{IJ}$ denotes the cofactor (i-e determinant of the
minor) of the $5\times 5$ Graham matrix. The equation
(\ref{eq:4s}) follows from eq. (\ref{eq:invmeas}) since the minor
$\L_{33}$ (resp. $\L_{44}$) is the Graham determinant of the
3-simplex $0124$ (resp. $0123$).

One of the main difference with the case of the $6j$ is the fact
that the angles $\t_{IJ}$ are not all independent. It is clear
from our parameterization that $\t_{34}$ is not needed : it is a
function of the 9 others angles. Indeed, from the original
integral expression, the 5 integration elements $g_I\in S^3$ can
be considered as $5$ unit vectors in $\real^4$. They are not
independent and the angles $\t_{IJ}$ can therefore be considered
as the dihedral angle of an Euclidean 4-simplex. This means that
the $5$ dimensional Graham matrix $G_{IJ}(\t)$ is degenerate,
$G(\t)= \det[G_{IJ}(\t)]=0$.

In order to get a symmetric expression of the integrant, let
$T(\t_{IJ})$ be a spherical 4-simplex whose dihedral angle are
$\t_{IJ}$, denote by  $G(\t)$ its  Graham determinant, and by
$\L_{IJ}$ its cofactor. It is clear that \be {\partial G(\t) \over
\partial \t_{IJ}}= -2\L_{IJ}(\t) \sin\t_{IJ}, \ee where the factor
2 is due to the symmetry of the Graham matrix. When the 4-simplex
is flat (i-e G=0) and non degenerate, the Graham matrix is of
corank one, which means that the cofactor matrix is of rank 1 \be
\L_{IJ}^2= \L_{II}\L_{JJ}. \ee This means that the denominator in
the integrand of (\ref{eq:4s}) is $|\partial_{34} G(\t)|/2$ and
the measure is simply
\begin{equation}
(\prod_{\begin{array}{c}\scriptstyle I<J,\\ \scriptstyle (I,J)
\neq (3,4)\end{array}}\!\!\! d\t_{IJ})\ \frac{1}{\sin\t_{34}
\sqrt{\L_{33}\L_{44}}} = (\prod_{I<J}d\t_{IJ})\ 2 \delta(G(\t)) .
\end{equation}
Using this measure in (\ref{eq:4s}) proves the theorem.}

\subsection{Asymptotics for the 10j symbol}
Following the lines of the method employed for the 6j symbol's
study, we split the integral $I$ into a sum $I=I^< +I^>$, where in
$I^>$ the angles $\t_{IJ}$ are restricted to be in $[\e,\pi-\e]$,
$\e>0$ small enough. The asymptotic of the integral  $I^>(N
l_{IJ})$ is governed by stationary phase. Given $l_{IJ}$ we denote
by $T^*(l_{IJ})$ the Euclidean 4-simplex such that the area of the
face opposite to the edge $(IJ)$ is given by $l_{IJ}$, and denote
$\T(l)$ the dihedral angles of this tetrahedron. If $l_{IJ}$ are
such that $T^*(l_{IJ})$ is not degenerate and $\e
<\mathrm{min}(\T(l),\pi-\T(l))$ then \be I^>(Nl_{IJ}) \sim {1\over
N^{9/2}} {\cos(N(l_{IJ} +1)\T_{IJ}) \over V(l)}, \ee where $V(l)$
has to be determined. This property was proven by Barret and
Williams \cite{BarW}. Since the argument of the  cosinus is the
Regge action, it was a strong evidence that spin foam models
constructed from the $10j$  give amplitude of 4d  gravity. Note
that if $T^*(l)$ is degenerate, then for $\epsilon>0$, there is no
stationary point in the action and $I^>(Nl_{IJ}) =o({1/
N^{9/2}})$.

However, we have to take into account the asymptotic behavior of
$I^<$. First, as for the $6j$, we find that the integration domain
of $I^<$ is split into 16 disconnected components. All this
components are equal and  related by the symmetry  transformation
$\t_{IJ} \to \sigma_I \sigma_J\t_{IJ}+(1- \sigma_I \sigma_J)\pi/2
$. With this symmetry, we can restrict the angles to be in
$[0,\e]$.
 According to our analysis of the $6j$ the asymptotic behavior of this integral is governed by the
singular contributions of the integrand, i-e points where
$\partial_{IJ}G(\t)=0$. The most singular points are indeed the
ones where $\t_{IJ}=0$. The behavior of the Graham determinant
around this point is given by
\begin{equation}
G(Su)\sim_{S \to 0} S^8 [4! V_{4s}(u)]^2
\end{equation}
where $V_{4s}(u)$ is the volume of the Euclidean 4-simplex whose
edge lengths are $u_{IJ}$. We can perform an analysis which is
almost identical to the one of section \ref{sec:asI<}, therefore
we do not repeat it. This analysis shows that
 the asymptotic of $I^<$ is given by the following contribution
\be I^>(Nl_{IJ}) \sim {16 \over N^2} {4\over
\pi^6}\int_{\cD_{\infty}} \prod_{I<J}\sin(l_{IJ}u_{IJ})
\delta([4!V_{4s}(u)]^2)\ [\prod_{I<J} du_{IJ}] \ee where the
integration is on $\cD_{\infty}$, the set of euclidian flat
4-simplices.

By an analysis similar to the one done in the previous section, we
can express the measure in term of a fully gauge fixed measure \be
{4\over \pi^6} \delta([4!V_{4s}(u)]^2)
(\prod_{IJ}du_{IJ})=[\prod_{I=0}^4 {d^3\vec{u}_{I}\over 2\pi^2}],
\ee where  $u_{IJ}= |\vec{u}_{J}-\vec{u}_{I}|$. The asymptotic
contribution of $I^<$ is therefore \be {16\over N^2}
\int_{(\real^3)^4} \prod_I \left[{d^3\vec{u}_{I}\over
2\pi^2}\right]\prod_{I<J}{\sin(l_{IJ}|\vec{u}_{J}-\vec{u}_{I}|)\over
|\vec{u}_{J}-\vec{u}_{I}|}
 \ee which is the
form proposed by Baez et al \cite{Baez1}.

\section{Discussion}\label{sec:disc}
As we mention in the introduction, given a triangulation of a
4-dimensional manifold, we can construct state sum models (usually
called Barrett-Crane models) where the amplitude associated with
the 4-simplex is given by the $10j$ symbols. The importance of
these models lies in the hope that they would give a way to
compute  transition amplitudes in 4d gravity \cite{BarC}. Also, it
was  found that the integral form of the $10j$ \cite{Barint} lead
to the interpretation of these weights as Feynman diagrams
\cite{FK}. This leads into the striking result that the sum over
all triangulation of Barret-crane amplitude could be interpreted
as a sum over Feynman graph of a non local field theory \cite{BO}.

However this model suffers from two diseases. First, the
amplitudes calculated in terms of $10j$ symbols are always
positive \cite{Baezpos} and second, as we have just seen, their
asymptotics is not dominated by the semi-classical Regge action
but it is dominated by degenerate configurations. Does this mean
that these model are therefore not good candidates for quantum
gravity amplitudes? In order to answer this, we have to recall why
the Barret-Crane models are suspected as good candidates for GR
amplitudes. The original derivation of the model in \cite{BarC}
did not make any reference to the Einstein-Hilbert action, however
there has been work since then linking the Barrett-Crane model
with a discretization of the path integral for Euclidean 4d
gravity \cite{LF,MR}. In these works it was argued that the
Barret-Crane model corresponds to a discretization of a constraint
$BF$ model introduced by Plebanski \cite{Pleb}, and it was shown
that the partition function of this model was related to the
gravity partition function if one excludes the degenerates metrics
from the path integral. However in \cite{MR} it was argued that
the degenerate sector is likely to dominate to path integral since
the number of degenerate configurations is likely to be more
important that the number of non degenerate one. A similar fact is
well known in statistical mechanics under the name `order by
disorder'\cite{Peter}. It is best illustrated in  the $XY$
antiferromagnetic model on a Kagome lattice. This model is
frustrated and posses a huge degeneracy (the  vacua states have a
macroscopic entropy). Since all the states have the same energy,
only the entropy of the fluctuation around these states could make
a difference between them. When the temperature is small, the
states which are selected are the soft modes states, (i-e the one
around which fluctuation in energy is quartic), whereas the hard
modes (the ones for which the fluctuation in energy is quadratic)
are suppressed. In Euclidean gravity we are working with the
Einstein-Hilbert action \be S=\int e^I \wedge e^J \wedge F^{KL}(A)
\e_{IJKL}, \ee where $e^I$ is the frame field $I$ are
$\mathrm{SO}(4)$ indices and $F^{IJ}(A)$ is the curvature of the
$\mathrm{SO}(4)$ spin connection. If we evaluate the action on
solutions of the equation of motion we find that $S=0$. So the
on-shell configurations are not distinguished by their action. By
a similar argument to the one used in statistical mechanics one
expect that the dominant configuration of the path integral are
given by the soft modes in the limit where $\hbar \to 0$. It is
clear that the fluctuations of the action around a configuration
where the metric (hence $e$) is non degenerate are purely
quadratic. The softest modes of $S$ is given by  $e=0$, and one
expect that fluctuations around this configuration dominates the
path integral in the classical limit if one include degenerate
metrics. Note that in $3$d the fluctuation around $e=0$ are
quadratic, so in this case, we do not expect the degenerate
contributions to
 dominate the non degenerate ones.

This argument suggests that the disease we have just find in the
asymptotic behavior of the 10j symbol is in fact a disease of the
gravity partition function if one includes degenerate metrics. The
cure for this disease is then clear, one has to avoid degenerates
contributions. How this can be implemented in the state sum model
is also clear. We have seen that given $\e>0$ the integral
defining the $10j$ symbol could be split int $I^< +I^>$ where
$I^<$ account for the dominant degenerate contributions, whereas
$I^>$ has a nice semi-classical oscillatory behavior. One can
therefore propose a model where, instead of taking the 10j symbol
as an amplitude for the 4-simplex, one takes the truncated 10j
symbol $I^>$. This cures at once the two diseases mentioned
earlier.

One of the key feature of the Barret-Crane weight was the fact
that it could be interpretated as a Feynman graph of a group field
theory. The modification of the weight we propose still preserve
this key property since $I^>$ can be written as a Feynman integral
with the propagator \be K_{\e,l}(g_1,g_2) = \chi_l(g_1g_2^{-1})
C_\e(g_1g_2^{-1}) \ee where $\chi_l$ denotes the character of the
representation of weight $l$ and $C_\e(e^X)$ is a cut-off function
which is $0$ if $|X|<\e$ and $1$ if $|X|>\e$. This let open the
possibility of writing the sum over triangulations of state sum
model constructed with a truncated 10j symbol as a Feynman graph
expansion. The drawback of the truncated model is the presence of
the cut-off $\e$ which still deserve a physical interpretation. We
hope to come back to these issues in the near future.

\vspace{5mm}\noindent{\bf Acknowledgement :} We would like to
thank M. Reisenberger for discussion. D. L. is supported by a
MENRT grant and by Eurodoc program from R\'egion Rh\^one-Alpes. L.
F. is supported by CNRS and an ACI-blanche grant.


\appendix
\section{Geometry of the Euclidean and  Lorentzian  flat tetrahedron}
\label{flat}

Let $\vec{e}_I \in \real^3 $, $ I=0,1,2,3$ be the vertices of a non degenerate tetrahedron,
 we denote by $\vec{A}_I$ the vectors normal to the face opposite to the vertex $I$.
They are defined as \be \vec{A}_I = {1\over 2}
(\vec{e}_J\wedge_\eta \vec{e}_K +\vec{e}_K\wedge_\eta \vec{e}_L+
\vec{e}_L\wedge_\eta \vec{e}_J) \ee where $I,J,K,L$ is an even
permutation of $0,1,2,3$, and \be (u \wedge_\eta v)^a = \eta^{ae}
\epsilon_{ebc}u^b v^c, \ee
 where $\e_{abc}$ is the totally antisymmetric tensor
with $\e_{123}=1$ and $\eta_{ab} =\eta^{ab}$ is the metric $+++$ in the Euclidean case
and $-++$ in the Lorentzian case.
In all the following we denote  the wedge product simply $\wedge$ and the scalar product with respect to $\eta$
by $u\cdot v$ without refereing explicitly to the dependence of the metric.
The vectors $\vec{A}_I $ clearly satisfy
\be\label{sum0}
\sum_I \vec{A}_I =0.
\ee
The edge vectors are given by $\vec{l}_{IJ}=\vec{e}_J- \vec{e}_I$, they satisfy with the area vectors a duality relation
\be
\vec{A}_i \cdot \vec{l}_{0j} = 6V \delta_{ij},
\ee
where $ V= (\vec{l}_{01}\wedge \vec{l}_{02})\cdot \vec{l}_{03} /6$ is the oriented volume of the tetrahedron.
Moreover if one uses the identity
$ (u \wedge v) \wedge w= \eta [(u\cdot w) v -(v\cdot w)u ]$, where $\eta$ is the signature of the metric, one gets
\be\label{wedge}
\vec{A}_I \wedge \vec{A}_J = \eta{3 V \over 2} \vec{l}_{KL}
\ee
where $I,J,K,L$ is an even permutation of $0,1,2,3$.

The Cayley matrix associated with the tetrahedron is defined as
\begin{equation}
C=
\left(\begin{array}{ccccc}
0        & l^2_{12} & l^2_{13} & l^2_{1}  & 1 \\
l^2_{12} & 0        & l^2_{23} & l^2_{2}  & 1 \\
l^2_{13} & l^2_{23} &    0     & l^2_{3}  & 1 \\
l^2_{1} & l^2_{2} & l^2_{3}    &     0    & 1 \\
1        & 1        & 1        &     1    & 0
\end{array}\right)
\end{equation}
where $l^2=\vec{l}\cdot \vec{l}$ and $l_{i}\equiv l_{0i}$.
This matrix encodes a lot about the geometry of the tetrahedron as  the following identities show
\beq
\det(C) &=& \eta 2^3 (6V)^2, \label{V} \\
\vec{A}_I \cdot \vec{A}_J &=& -{\eta \over 16 } \D_{IJ},
\label{co} \eeq where $\D_{IJ}$ is the cofactor matrix  of the
Cayley matrix.

The first relation can be proven as follows,
we have the identity
\begin{equation}
\det(\vec{l}_i\cdot \vec{l}_j)=\det(\vec{l}_i^a\eta_{ab}
\vec{l}_i^b) =\det(\vec{l}_i^a)^2 \eta = \eta (6V)^2
\label{volumetetra}
\end{equation}
Using the relation.
\begin{equation}
\vec{l}_i\cdot \vec{l}_j= \frac{l_i^2+l_j^2-l_{ij}^2}{2}
\end{equation}
The square of the volume is rewritten as
\begin{equation}
2^3 6^2 V^2 = \left|\begin{array}{ccc}
2l_1^2                & -l_{12}^2+l_1^2+l_2^2 & -l_{13}^2+l_1^2+l_3^2 \\
-l_{12}^2+l_1^2+l_2^2 & 2l_2^2                & -l_{23}^2+l_2^2+l_3^2 \\
-l_{13}^2+l_1^2+l_3^2 & -l_{23}^2+l_2^2+l_3^2 & 2l_3^2
\end{array}\right|
\end{equation}
This $3\times 3$ determinant is easily seen to be the Cayley determinant.
This follows from substracting line 4 to lines 1,2 and 3, then row 4 to rows 1,2 and 3
in the Cayley determinant to finally obtains
\begin{equation}
 \left|\begin{array}{ccccc}
-2l_1^2                & l_{12}^2-l_1^2-l_2^2 & l_{13}^2-l_1^2-l_3^2 & l_1^2 & 0\\
l_{12}^2-l_1^2-l_2^2 & -2l_2^2                & l_{23}^2-l_2^2-l_3^2 & l_2^2 & 0\\
l_{13}^2-l_1^2-l_3^2 & l_{23}^2-l_2^2-l_3^2 & -2l_3^2                & l_3^2 & 0 \\
l_1^2                 & l_2^2                 & l_3^2                 &    0  & 1 \\
0                     & 0                     & 0                     &    1  & 0
\end{array}\right|
\end{equation}
which reduces to the $3\times 3$ determinant.

The relation
(\ref{co}) is proven by direct computation and the use of the identity
\be
(u\wedge v)\cdot (w\wedge x) = \eta[(u\cdot w)(v\cdot x) -(u\cdot x)(v\cdot w)]
\ee

Lets denote $A_I=\sqrt{|\vec{A}_I\cdot \vec{A_I}|}$ the area of
the face $I$, lets introduce the normal unit vector $\vn_I$,
$\vec{A}_I = A_I \vn_I$ and lets denote $G_{IJ}= \vn_I\cdot \vn_J$
the Graham matrix. The relation (\ref{sum0}) implies that
\begin{equation}
\sum_J A_J G_{IJ}=0. \label{zeroG}
\end{equation}
From relation (\ref{co}), we get that for $I,J=1..4$, the cofactor
$\D_{IJ}$ are related to the elements of the Graham matrix
\begin{equation}
\D_{IJ}=-\n 16 A_I A_J G_{IJ}
\end{equation}
A consequence of eq.\ref{zeroG} is that
\begin{equation}
\sum_{I=1}^4 \D_{IJ}=0
\end{equation}
If the tetrahedron is Euclidean $G_{IJ}= \cos\T_{IJ}$, and the
relation (\ref{wedge}) imply \be A_I A_J \sin\T_{IJ}=\frac{3}{2}
|V| l_{KL}. \ee In this case the Cayley determinant is positive.

If the tetrahedron is Lorentzian and time-like (i-e all its edge
are time-like $\vec{l}_{IJ}\cdot\vec{l}_{IJ}=-l_{IJ}^2$) then the
normal vectors are all space-like and $G_{IJ}= \cos\T_{IJ}$  and
\be A_I A_J \sin \T_{IJ}=\frac{3}{2} |V| (\pm) l_{KL}. \ee where
the sign in the RHS is $+$ (resp. $-$) if the edge $\vec{l}_{KL}$
is future (resp. past) pointing. In this case the Cayley
determinant is negative. Conversely given a matrix $G_{IJ}=
\cos\T_{IJ}$ which is degenerate (of corank 1) one can associate
uniquely to it a pair of Euclidean tetrahedra or a unique
Lorentzian tetrahedron.

If the tetrahedron is Lorentzian and space-like (i-e all its edge
are space-like $\vec{l}_{IJ}\cdot\vec{l}_{IJ}=\rho_{IJ}^2$) then
the normal vectors are all timelike and $G_{IJ}= \nu_{IJ}\cosh
T_{IJ}$, $T >0$, $\nu_{IJ}=\pm 1$ and $\nu_{II}=-1$. $\v_{IJ}$ is
the sign of the scalar product $(n_I \cdot n_J)$, depending on
whether $n_I$ and $n_J$ lie in the same part of the light-cone.
The $\v_{IJ}$ can thus always be written as $\v_{IJ}=-\a_{I}\a_J$
with $\a_I=\pm 1$. We call the pair $(\nu_{IJ}, T_{IJ})$ where $T
>0$ $\nu_{IJ}=\pm1$ the dihedral angle of the edge $(IJ)$. The
relation(\ref{wedge}) implies \be A_I A_J \sinh T_{IJ}=\frac{3}{2}
|V| \rho_{KL}. \ee

\section{Some facts about $\SL(2,\real)$}\label{sl}
We denote by $T_{il^+}$, $l \in \Nat$ the positive discrete series
of weight $l$, $T_{-il^-}$, $l \in \Nat$ the negative discrete
series of weight $l$, and $T_{\rho}$ $\rho \in \real^+$ the
principal series of weight $\rho$. In this paper we consider only
the representations of $\SL(2,\real)$ which extend to
representations of $\SO(2,1)$, this essentially means that $l \in
2\Nat$. It is well known that the continuous representation are
conjugated to itself, whereas the positive discrete series is
conjugated to the negative discrete series. So if one change the
orientation of one edge this amounts to change the label $\l$ into
its complex conjugate $\bar{\l}$. Let $(\l_1,\l_2,\l_3)$ be a
triple of representations meeting at a vertex of the oriented
tetrahedron, modulo a change of orientation we can suppose that
all the edges are incoming at the vertex. Such a triplet is always
admissible if there is at least one continuous representation. If
there is no continuous representation the admissible triplets (for
the incoming orientation) are

$(l_{1}^{+}, l_{2}^{+}, l_{3}^{+})$ with  $l_{3} > l_{1}+l_{2}$
and $l_{1}+l_{2} +l_{3} \in 2 \Z $;

$(l_{1}^{-}, l_{2}^{-}, l_{3}^{-})$ with  $l_{3} > l_{1}+l_{2}$
and $l_{1}+l_{2} +l_{3} \in 2 \Z $.

Let $g$ be
an element of $ \SL(2,\real)$. We say that  $g$ is
elliptic if it is conjugated to the matrix
\be
k_\t=
\left(
\begin{array}{cc}
\cos \t & \sin \t \\
-\sin\t & \cos \t
\end{array}\right).
\ee
$g$ is said to be
hyperbolic if it is conjugated to the matrix
\be
\pm a_t=
 \pm \left(
\begin{array}{cc}
e^t & 0 \\
0 &  e^{-t}
\end{array}\right).
\ee Note that $k_\t$ and $k_{-\t}$ are not conjugated to each
other in $\SL(2,\real)$, whereas $a_t$ and $a_{-t}$ are. The
characters of the positive discrete series are given by
\cite{Knapp} \be \chi_{il^+}(g)=  - \frac{1}{2i}{e^{i(l-1)\t}
\over \sin\t} \ee if $g$ is conjugated to $k_\t$ and \be
\chi_{il^+}(g)= \frac{e^{-(l-1)|t|}}{2|\sinh t|} \ee if $g$ is
conjugated to $\pm a_t$ $t>0$. The characters of the negative
discrete series are conjugate to the positive discrete series
characters \be \chi_{il^+}(g)=\bar{\chi}_{-il^-}(g) \ee The
characters of the principal series are given by \be
\chi_{\rho}(g)=0 \ee if $g$ is conjugated to $k_\t$ and \be
\chi_{\rho}(g)= \frac{\cos\rho t}{|\sinh t|} \ee if $g$ is
conjugated to $\pm a_t$ $t>0$.

Since the representations are infinite dimensional the characters
of these representations are not function but distributions. The
previous expressions should be consider as a particular
representative of the character distribution in terms of a locally
integrable function. One has however the freedom to change the
representative function as long as they differ from the previous
one on a set of measure $0$. It turns out that we effectively need
to take an other representative function in order to make sense of
the integral (\ref{6jLor}), since the integral admits
singularities around $g_I =g_J$. We propose the following
modification. Lets consider $\alpha>0$ and lets define
$C_\alpha(g)$ to be equal to $1/({e^t -e^{-t}})$ if  $g$ is
conjugated to $\pm a_t$, $0<t<\alpha$ and to be equal to $0$
otherwise. For the continuous series $\rho$ lets consider the
following distribution \be <{\chi}^\prime_\rho|f> = lim_{\alpha
\to 0}\int_G (\chi_\rho(g) - C_\alpha(g))f(g). \ee If $f$ is
regular at the identity then it is clear that
$<{\chi}^\prime_\rho|f>=<{\chi}_\rho|f>$ so ${\chi}^\prime_\rho$
define the same distribution as $\chi_\rho$. Analogously we define
${\chi}^\prime_{\pm l^\pm}$ by substracting the divergence at $0$.
And we use these characters in the definition of the integral
(\ref{6jLor}). It is easy to see that with such a representative
of the character the integral is convergent.

\end{document}